\documentclass{article}
\AtBeginDocument{%
  \providecommand\BibTeX{{%
    \normalfont B\kern-0.5em{\scshape i\kern-0.25em b}\kern-0.8em\TeX}}}

\usepackage{xspace}
\usepackage{import}
\usepackage{amssymb}
\usepackage{hyperref}
\usepackage{url}
\usepackage{natbib}
\usepackage{graphicx}
\usepackage{amsmath}
\usepackage{subfig}
\usepackage{float}
\usepackage{authblk}
\usepackage{capt-of}
\usepackage{wrapfig}
\usepackage{xcolor,colortbl}

\title{YouTube over Google's QUIC vs Internet Middleboxes: A Tug of War between Protocol Sustainability and Application QoE}
\author[1]{Sapna Chaudhary}
\author[1]{Prince Sachdeva}
\author[3]{Abhijit Mondal}
\author[1]{Mukulika Maity}
\author[2]{Sandip Chakraborty}

\affil[1]{Department of Computer Science and Engineering,IIIT Delhi, India}
\affil[2]{Department of Computer Science and Engineering,IIT Kharagpur, India}
\affil[3]{VDX.TV, Noida, India}
\date{}                     
\begin{document}


\maketitle

\begin{abstract}
Middleboxes such as web proxies, firewalls, etc. are widely deployed in today's network infrastructure. As a result, most protocols need to adapt their behavior to co-exist. One of the most commonly used transport protocols, QUIC, adapts to such middleboxes by falling back to TCP, where they block it. In this paper, we argue that the blind fallback behavior of QUIC, i.e., not distinguishing between failures caused by middleboxes and that caused by network congestion, hugely impacts the performance of QUIC. For this, we focus on YouTube video streaming and conduct a measurement study by utilizing production endpoints of YouTube by enabling TCP and QUIC at a time. In total, we collect over $2600$ streaming hours of data over various bandwidth patterns, from $5$ different geographical locations and various video genres. To our surprise, we observe that the legacy setup (TCP) either outperforms or performs the same as the QUIC-enabled browser for more than $60$\% of cases. We see that our observation is consistent across individual QoE parameters, bandwidth patterns, locations, and videos. Next, we conduct a deep-dive analysis to discover the root cause behind such behavior. We find a good correlation ($0.3-0.7$) between fallback and QoE drop events, i.e., quality drop and re-buffering or stalling. We further perform Granger causal analysis and find that fallback Granger causes either quality drop or stalling for $70$\% of the QUIC-enabled sessions. We believe our study will help designers revisit the decision to enable fallback in QUIC and distinguish between the packet drops caused by middleboxes and network congestion.
\end{abstract}

\section{Introduction}
Transmission Control Protocol (TCP) has controlled the Internet for over three decades; however, in the past few years, \textit{Quick UDP Internet Connection} (QUIC), developed initially and publicly announced by Google in 2013, has slowly become the predominant replacement of TCP. HTTP/3, although an Internet-Draft as of now, uses QUIC at its core, while major Internet content providers, like Facebook, Cloudflare, Apple, Akamai, etc., have already migrated towards supporting their apps over HTTP/3. Many existing studies have argued that QUIC is advantageous over TCP for heavy-tailed latency characteristics of the Internet~\cite{langley2017quic} and thus is more suitable to support consistent quality of experience (QoE) for applications like video streaming. However, there also exist works that counter-argue such facts and show that there are instances when QUIC suffers in comparison with TCP~\cite{carlucci2015http, cook2017quic, kakhki2017taking, nepomuceno2018quic, ruth2019perceiving,yu2017quic}, although they failed to point out the root cause behind such observations clearly. This paper performs a thorough analysis of QUIC's behavior over an experimental setup that streamed more than $2600$ hours of YouTube videos over both TCP and QUIC over different network setups and from various geographical locations. The analysis reveals some interesting insights about QUIC's long and short-term protocol behaviors that can answer a few open questions about QUIC's in-the-wild performance over the Internet. 

As QUIC has been developed as a replacement for TCP, it fundamentally uses UDP at its core to transfer data segments between two end hosts. Interestingly, the current Internet is full of middleboxes~\cite{middlebox-more}, like firewalls, proxies, load balancers, intrusion detection systems, etc. Such middleboxes typically do not prefer UDP as the transport protocol for the apparent reason of network security, and thus block the UDP connections~\cite{ietf}. To alleviate this problem, QUIC adapts its behavior where the application client races a TCP connection with QUIC; whichever finishes first (can successfully establish the connection) gets used to serving the application requests. In this case, the failed QUIC connection is marked as broken for a timeout duration or as decided by an exponential backoff mechanism, and the subsequent application requests fall back to TCP until a successful QUIC connection is established again~\cite{langley2017quic, quic-fallback}. 

While the above fallback mechanism helps an application continue a connection with the remote host even when the middleboxes block UDP, it might have a potential side effect causing an unexpected impact on an application's performance. The fallback mechanism does not explicitly check whether the QUIC connection failure is due to the presence of a middlebox blocking the underlying UDP segments. It is highly possible to have a momentary connection failure due to poor end-to-end network connectivity between the two remote hosts. Although a QUIC client uses 1-RTT or 0-RTT connection establishment with the remote host, the connection establishment packets (CHLO, REJ, etc.) are over unreliable UDP. Consequently, QUIC connection establishment uses one additional level of redirection (application $\rightarrow$ transport $\rightarrow$ network) compared to TCP connection establishment packets (transport $\rightarrow$ network). Consequently, the QUIC connection establishment packets may fail over a lossy heavy-tailed network, but the TCP connection establishment packets are successfully delivered. Therefore, even in the absence of a middlebox or when the middlebox allows UDP, QUIC may fall back to TCP when the end-to-end network connectivity is poor. With this motivation, we ask the following questions in this paper.
\begin{enumerate}
    \item \textit{How frequently do we observe a fallback over a QUIC-enabled stream from a QUIC-supported browser?} This analysis can shed light on whether the majority of fallbacks are related to middleboxes blocking the UDP or can even happen when there are no middleboxes. Also, we tend to see the network conditions that can trigger such fallbacks. 
    \item \textit{How does QUIC fare compared to TCP in terms of long-term performance of a typical application like video streaming}\footnote{We consider video streaming in this paper, as the streaming traffic mostly dominates the Internet and has stringent QoE requirements.} \textit{over a heavy-tailed network?} The answer to this question can help us to understand the sustainability of QUIC in the long term over the current Internet. Indeed, the QUIC developers~\cite{langley2017quic} and several other works~\cite{kakhki2017taking, quicd, chrome-blog, cloudfare-blog, medium-blog} have shown the superiority of QUIC compared to TCP in terms of supporting better QoE with less stall for video streaming. 
    \item \textit{Does a video streaming application's short-term or instantaneous performance get impacted when QUIC is enabled over the browser?} We presume that protocol fallbacks may not affect QUIC performance in the long run; still, it may create intermediate glitches in the application QoE and thus jeopardize the protocol performance compared to TCP for applications like video streaming. 
    \item \textit{How severely does the middlebox-adaptive behavior of QUIC (fallback to TCP) impact the QoE of a video streaming application?} Specifically, this paper tries to analyze whether the fallback is one of the primary reasons behind QUIC's poor performance over the heavy-tailed network, as pointed out in a few previous literature~\cite{nepomuceno2018quic,yu2017quic}.  
\end{enumerate}

We set up a testbed to capture the browser traffic streamed from the YouTube server with two different setups -- (1) enabling QUIC and (2) disabling QUIC in the browser setup. We create a list of $46$ different videos (mean video duration of $3600$ secs) from various genres such as news, entertainment shows, education, talk shows, comedy shows, etc.; our collected dataset contains video traces for more than $2600$ streaming hours over $5$ different geographical regions. We collect both the application and network level logs to analyze the video streaming QoE as well as its dependency and relationship with the underlying traffic patterns.

\subsection{Contributions}
In contrast to the existing works, our contributions in this paper are as follows.\\

\noindent\textbf{(1) Analyzing QUIC from an Application's Viewpoint:} This paper analyzes the performance of an application when QUIC is enabled in the browser as the recommended transport protocol over the Internet. This provides a more realistic view of the protocol's performance over the real Internet with its own hitches and glitches for the end-to-end connectivity. We observe that a traffic stream from a QUIC-enabled browser carries a significant amount of TCP traffic (even sometimes more than the total volume of QUIC traffic) when the network bandwidth is low. We thoroughly analyze how this protocol switching (or TCP fallback) impacts the performance of YouTube video streaming.\\

\noindent\textbf{(2) Characterizing the Race between QUIC and TCP:} To the best of our knowledge, this is the first work that analyzes the different scenarios when a QUIC-enabled browser transmits application data over TCP traffic rather than QUIC traffic, even if there are no middleboxes to block the QUIC (UDP) traffic. We observe that TCP traffic is highly likely to win the race over a low bandwidth and lossy network.\\

\noindent\textbf{(3) Impact of TCP Fallback on YouTube Streaming Performance:} With $2046$ video sessions combing more than $2600$ streaming hours over different bandwidth patterns across various geographical regions, we analyze how a QUIC-enabled browser impact the performance of YouTube streaming QoE (in terms of average playback bitrate and video stalling) in comparison with a legacy browser that uses pure TCP traffic to stream the video. We observe that enabling QUIC in the browser does not help improve the QoE continually; instead, YouTube suffers over a QUIC-enabled browser when the network bandwidth is low. We perform statistical testing, i.e., Welch's \textit{t}-test~\cite{t-test} and observed that a legacy browser (with pure TCP) either provides better QoE or the same QoE as a QUIC-enabled browser for $61\%$ cases. Interestingly, we found out that TCP traffic dominates over QUIC traffic under low bandwidth, even if the browser enables QUIC for data delivery.\\

\noindent\textbf{(4) Correlation and Causality Testing for Root-cause Analysis:} We perform thorough statistical analysis, i.e., correlation and causal tests, to check if fallbacks are the root causes behind sub-optimal application QoE of QUIC-enabled streams. We perform these analyses only for the QUIC-enabled sessions where pure TCP provided better QoE. We find a good correlation, i.e., the median is $0.55$ between fallback and quality drop/stalling. Next, we perform causality analysis using Granger causal model~\cite{Gc}. We observe that fallback Granger causes a quality drop in nearly $20$\% video sessions and stalling in nearly $60$\% video sessions. In total, fallback Granger causes either quality drop or stalling for $70\%$ video sessions.\\

We believe that our work is the first step towards uncovering the issues created by the fallback behavior of QUIC. This paper suggests that QUIC designers should take a deeper look at the failures caused by middleboxes and network congestion and adapt accordingly.

\section{BACKGROUND AND RELATED WORK}
\label{background-related}
This section first covers a short overview of QUIC, and then it highlights various related works on QUIC performance analysis in the wild.

\subsection{Quick UDP Internet Connection (QUIC)}
\begin{wrapfigure}{r}{0.4\textwidth}
\includegraphics[width=0.4\textwidth]{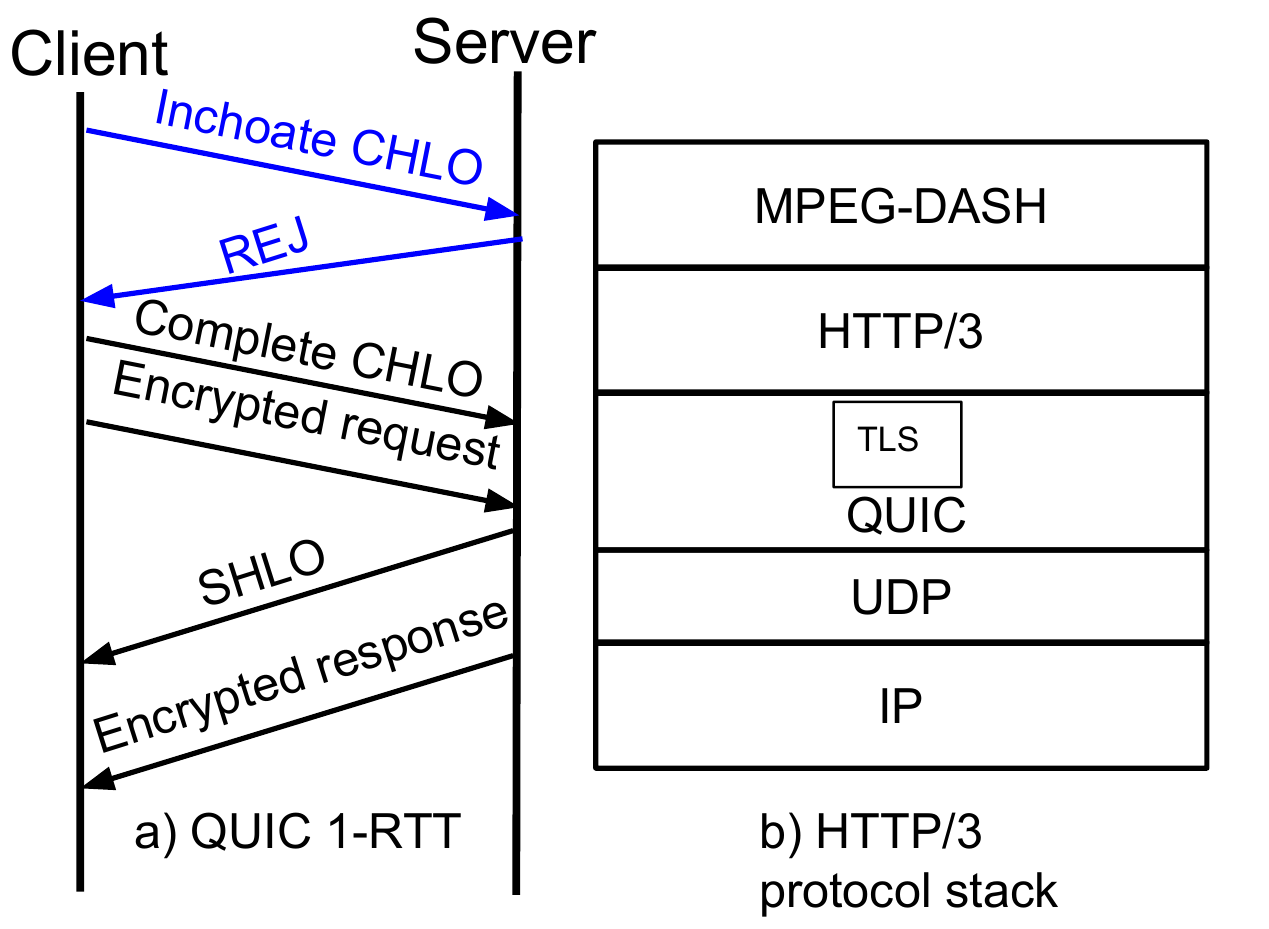}
\caption{(a) Timeline of QUIC's 1-RTT and 0-RTT handshake~\cite{langley2017quic} (b) ABR Streaming over HTTP/3 Protocol Stack with QUIC}
\label{quic-conn}
\end{wrapfigure}

Google first introduced QUIC in 2013; its development at the application layer makes it easier to modify, and so, since then, it has undergone several rapid growths. In 2012, Google introduced the SPDY protocol that can multiplex more than one HTTP request/response into one TCP connection to solve the TLS handshaking delay and connection establishment time for every HTTP request and response. But in the case of TCP, if even a single packet gets lost in that TCP stream due to in-order packet delivery, all the HTTP streams get blocked with SPDY, called the Head of Line (HoL) blocking problem. To solve the HoL problem, Google introduced a new protocol called QUIC. It is a connection-oriented protocol, and by using UDP, it can establish multiplexed connections between two endpoints and thus solves the HoL blocking problem of TCP by using UDP as transport protocol underneath. Fig.~\ref{quic-conn}(b) shows the protocol stack for HTTP/3 utilizing QUIC. It has improved the user Quality of Experience (QoE) by enhancing the QoE of HTTPS traffic and reducing the page load time. It has improved congestion control over multiplexed application flows and efficient recovery from the losses because of better estimation of RTTs. Also, it uses TLS 1.3 to encrypt the payload. QUIC combines the TLS 1.3 handshake with the typical handshake of TCP. 

Fig.~\ref{quic-conn}(a) shows how QUIC establishes a 1-RTT connection by exchanging a CHLO (client hello), then the server sends a REJ (Reject). In the case of 0-RTT, this exchange is not required. The client directly sends complete CHLO and an encrypted handshake request. The server sends hello and an encrypted handshake response. At this point, they can start the data communication. At a browser level, Google's Chrome races a TCP connection but prefers QUIC by delaying the TCP one by $300$ ms~\cite{langley2017quic,quic-fallback}. Whichever protocol wins the race gets used for that request. If QUIC's handshake fails either due to blocking of QUIC or QUIC's handshake packet being larger than the Maximum Transmission Unit (MTU), QUIC is marked as broken for a duration determined by an exponential backoff. The request then falls back to TCP.

\subsection{Adaptive Bitrate Streaming over HTTP/3} 
Modern adaptive bitrate (ABR) streaming, including YouTube, uses the protocol \textit{Dynamic Adaptive Streaming over HTTP} (DASH), also known as MPEG-DASH \cite{mondal2017candid}. DASH is a streaming standard that adapts the bitrate over the Internet based on the client's network conditions for streaming the video on the client-side. The video content delivered from the HTTP web server is broken down into smaller segments, and each segment is encoded into different bitrates at the server-side. The segments are requested and downloaded based on network conditions and stored in the receiver's buffer. DASH is a standard and not a protocol, and hence every video streaming application can have its own DASH implementation. 

As shown in Fig.~\ref{quic-conn}(b), the DASH client works on top of HTTP/3, and the DASH parameters are embedded within HTTP Request and HTTP Response messages. A DASH client decides the playback bitrate depending on the current network conditions; for this purpose, it uses a bitrate adaptation algorithm like BOLA~\cite{spiteri2020bola}, Pensieve~\cite{mao2017neural}, Oboe~\cite{akhtar2018oboe}, Fugu~\cite{yan2020learning}, etc. It can be noted that YouTube uses its own closed-source bitrate adaptation algorithm, although works like~\cite{krishnappa2013dashing,mondal2017candid,gutterman2019requet} have tried to reverse-engineer the YouTube ABR algorithm. Once a DASH client decides about the playback bitrate, depending on its local buffer state, it requests for the next video segment through an HTTP Request message that embeds the segment details and the required bitrate information (specified through a parameter called \textit{itag}). The DASH server then transfers the segment data using an HTTP response message. 

\subsection{Related Work}
We discuss the prior works \cite{ruth2019perceiving, nepomuceno2018quic, cook2017quic, kharat2018quic, nguyen2020scalable, wang2018implementation, bakri2015http, cui2017innovating, gutterman2019requet, arisu2018quickly, bhat2018improving, xu2020csi, madariaga2020analyzing, manzoor2020performance, aoki2021performance, moulay2021experimental, dissecting, qi2020performance, khan2020qoe} related to this paper from two broad perspectives: (a) QUIC and the DASH, (b) existing benchmarks on  QUIC vs TCP.\\

\textbf{QUIC Protocol and DASH:} Various recent works have analyzed the performance of adaptive bitrate streaming, particularly DASH, over QUIC and HTTP/3. Bhat \textit{et al.}~\cite{bhat2017not} studied the potential of using DASH over QUIC in place of DASH over TCP. They used average bitrate, quality switches, re-buffering ratio as the metrics to compare the performance of DASH over QUIC vs. the same over TCP. Interestingly, they found out DASH over QUIC does not provide much benefits compared to TCP; in fact, it degrades the playback bitrate at times. Mondal \textit{et al.} \cite{mondal2017candid} explored the implementation of YouTube's DASH client. They collected YouTube traces of $500$ videos using HTTP archives (HAR). They observed that (a) the distribution of buffer size at the client has a role in video quality adaptation, (b) whenever there is a drop in bandwidth, YouTube varies the segment length first before degrading the video quality. Further, they show that this leads to higher data wastage. Finally, they came up with an analytical model to predict data consumption of YouTube with a given initial quality and network conditions. In a follow-up work~\cite{mondal2020does}, the authors have shown that the stream multiplexing over the same UDP buffer creates a problem for DASH video and audio channels to synchronize, thus resulting in poor performance. In~\cite{engelbart2021congestion}, the authors have shown that QUIC congestion control needs to be tuned for supporting QoE for multimedia traffic. In the same line, a few other works~\cite{palmer2018quic,arisu2018quickly,van2018empirical,perkins2018real,herbots2020cross,lorenzi2021days} in the literature has also adopted the QUIC protocol to support better streaming performance; however, these works have not fundamentally looked within the interplay between DASH and QUIC instead used QUIC as a service.\\

\textbf{Performance comparison between QUIC and TCP:} Several studies have been conducted that aim to compare the performance QUIC with that of TCP~\cite{langley2017quic, kakhki2017taking, quicker, palmer2018quic, wolsing2019performance, seufert2019quic, biswal2016does, dissecting, carlucci2015http, das2014evaluation, megyesi2016quick, yu2017quic, ruth2019perceiving, nepomuceno2018quic, cook2017quic, chrome-blog, fb-blog, iyengar2018moving, saif2020early, kharat2018quic, nguyen2020scalable}. We compare the major related works in Table.~\ref{tab:related} and discuss some of them that are more recent and relevant to our study. In the seminal paper from Google on QUIC, Langley \textit{et al.}~\cite{langley2017quic} discussed the development and design of the QUIC protocol. They highlighted that QUIC reduces handshake delay, head-of-line blocking delay, and the dependency on network operators and middlebox vendors to deploy any change or modification in transport. They observed that QUIC reduces Google search time as $8$\% on desktop and $3.6$\% on mobile. It further decreased re-buffering rates on YouTube by $18\%$ on desktop and $15$\% on mobile. The authors discussed the fallback behavior of QUIC. They highlighted how $1$-bit change in the public field of QUIC resulted in some middleboxes allowing a few initial packets of QUIC but then subsequently blocking it. Therefore, QUIC packets face a destiny of black-hole. The authors realized the pain points of sustainability of a protocol across various middleboxes through ``\textit{deployment impossibility cycle}". Kakhki \textit{et al.} \cite{kakhki2017taking} conducted an in-depth analysis of QUIC and compared it with TCP. The authors focused mostly on web page load times and briefly on YouTube-based video streaming. They used a state machine diagram to analyze the states of the QUIC protocol. They make various interesting observations; we list some of them relevant to our study. (a) In a desktop environment, QUIC outperforms TCP for almost every case due to 0-RTT connection establishment and the ability to recover from losses quickly. (b) QUIC outperforms TCP in scenarios with fluctuating bandwidth due to its capability of eliminating ACK ambiguity, (c) QUIC either performs the same as TCP for video streaming or outperforms in the case of high-resolution video. This study made us believe we should expect a similar or better performance from QUIC than TCP for YouTube video streaming; however, the authors have not explicitly considered or mentioned the impact of fallbacks on application performance. In \cite{marx2020same}, the authors have compared and analysed 15 implementations of IETF QUIC using their tool \texttt{qlog} and \texttt{qvis}, where \texttt{qlog} is a logging format, and \texttt{qvis} is a tool that gathers the files of \texttt{qlog} and provides better visualization of the protocol for developing and debugging the QUIC protocol. They found huge heterogeneity between the QUIC implementations that need to be thoroughly analyzed.

\begin{table}[t]
    \caption{Existing works on QUIC vs. TCP Performance (TCP$>$QUIC indicates that the authors have shown TCP performs better than QUIC, QUIC$\simeq$TCP indicates QUIC performs similar to TCP; MBox: Middlebox)}
    \label{tab:related}
    \scriptsize
    \centering
    \resizebox{12cm}{!}{
    \begin{tabular}{|p{0.8cm}|p{2cm}|p{1.2cm}|p{2.5cm}|p{1cm}|p{1cm}|p{1cm}|p{1cm}|}
    \hline
      \textbf{Work} &  \textbf{Type of Application Traffic} & \textbf{Bandwidth (Mbps)} & \textbf{Observation} & \textbf{Root-cause Analysis} & \textbf{Updates to QUIC} &
      \textbf{Impact of MBoxes} &
      \textbf{Impact of Fallback} \\
      \hline
      \cite{langley2017quic} & Web Browsing and Youtube streaming & - & Web \& Youtube: QUIC>TCP & Yes & Yes & Yes & Yes\\
      \hline
      \cite{kakhki2017taking} &  Mostly web browsing \& limited video streaming & Fixed five levels & Web: QUIC$>$TCP. Video: QUIC$\simeq$TCP mostly, QUIC$>$TCP high res. & Yes & No & No & No\\
    \hline
    \cite{quicker} & Youtube video streaming & Fixed two levels & QUIC$\simeq$TCP & No & No & No & No\\
    \hline
    \cite{palmer2018quic} & Video streaming & Fixed & QUIC$\simeq$TCP & No & Yes & No & No\\
    \hline
    \cite{wolsing2019performance} & Web Browsing & Dynamic & QUIC$>$TCP & No & No & No & No\\
    \hline
    \cite{dissecting} & Web Browsing & Fixed & QUIC$\simeq$TCP & Yes & No & No & No\\
\hline 
    \rowcolor{cyan}
    \textbf{{Our Work}} & {YouTube video streaming} & {Dynamic} & {TCP$>$QUIC for video streaming over low bandwidth network} & {Yes} & {No} & {Yes} & {Yes}\\
\hline
    \end{tabular}
    }
\end{table}

Recently, Seufert \textit{et al.}~\cite{quicker} conducted a measurement study of TCP and QUIC's performance over $900$ YouTube video streaming sessions. They used the following metrics like quality of experience:  initial playback delay, video quality, and stalling. They compared the results of TCP and QUIC statistically and did not observe any evidence of QoE improvement over TCP. Their results match our observations; however, the authors did not conduct any root cause analysis for this behavior. Interestingly, the authors collected over $900$ videos but ultimately used only $500$ videos. They claim that they had inspected these flows to ensure that the videos were either streamed via TCP or QUIC. We believe due to fallback, some of the QUIC streams were discarded as they contained a significant amount of TCP packets. 
Palmer \textit{et al.} \cite{palmer2018quic} conducted measurement and claimed that neither TCP nor QUIC is well suited for video streaming. They quote the reliability feature of the QUIC protocol and note that not all types of frames need reliability. Specifically, I-Frames need to be transferred reliably but not P-Frames or B-Frames. But, QUIC does not support partial reliability. Hence, they proposed an extended version of QUIC called ClipStream that offers reliability only on selected frames, i.e., I-Frames. They develop a prototype and observe that the proposed approach outperforms QUIC and TCP in QoE. Such an approach is complementary to our study. Wolsing \textit{et al.} \cite{wolsing2019performance} also focuses on the comparison between QUIC and TCP. They point out that most research compared optimized QUIC with classical TCP stack, which is unfair. Hence, they fine-tune TCP's parameters and then perform the comparison. The authors observe that the QUIC still performs better than TCP, but the gap narrowed down. Yu \textit{et al.}~\cite{dissecting} focused on a comparison of TCP and QUIC in the production environment of Google, Facebook, and Cloudflare. The authors highlight that the benefits of QUIC, i.e., reduced RTT in connection establishment, avoiding head-of-line blocking, etc., can only be realized with a correct focus on implementation and configuration. The authors point out that discrepancies in performance between TCP and QUIC are mostly due to implementation details which are not specific inherent properties of the QUIC protocol. However, this can be difficult and time-consuming to deal with edge cases and implement congestion control from scratch. 

\section{Experimental Setup and Dataset}
\label{sec:expt-setup}
\begin{wrapfigure}{r}{0.5\textwidth}
\centering 
    \includegraphics[width=0.5\textwidth]{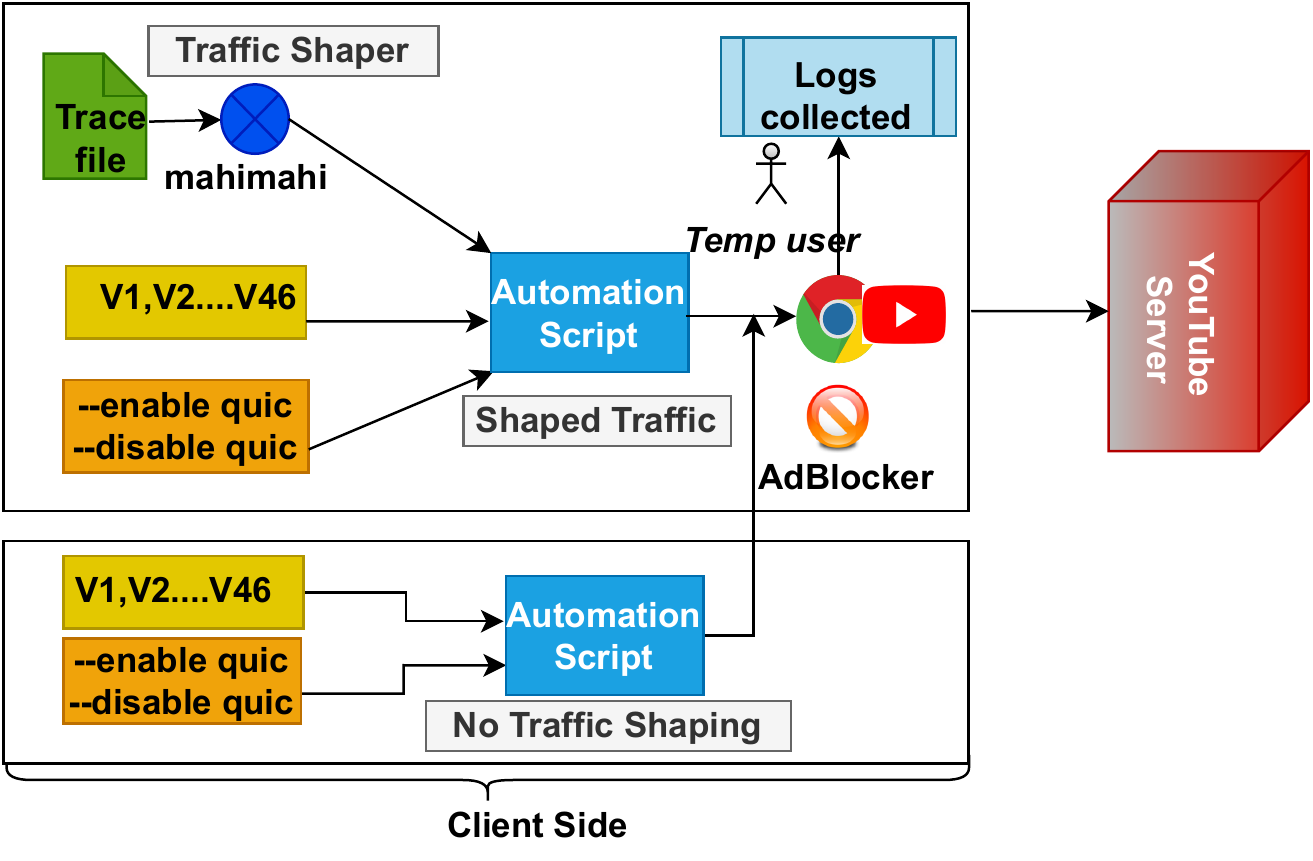}
    \caption{Experimental setup to stream YouTube videos and collect logs}
    \label{exp-setup}
\end{wrapfigure}
This section describes our experimental setup in detail. For the experiment, we created a list of $46$ videos selected from YouTube, each with duration $2400-6000$ sec. 
The details are shown in Table~\ref{video-details}.
Fig.~\ref{exp-setup} shows our experimental setup, which contains a pipeline for streaming the YouTube videos through a traffic shaper (Mahimahi) that controls the Internet bandwidth based on some predefined patterns. We played the videos across different geographical locations, namely Delhi, Bangalore, New York, Germany, and Singapore. One of the testbeds is setup inside our campus premises and the rest are setup using Digital Ocean machines, each having $x86\_64$ architecture with the processor running at $2494.136$ MHz frequency, $4$ GB RAM, and with Ubuntu 18.04 operating system. The details follow.




%

%
\subsection{Browser Setup} 
We stream YouTube videos from the Google Chrome browser. Google Chrome supports both the protocol stacks to enable or disable QUIC protocol for a specific YouTube streaming. We set the QUIC protocol by setting the flag \texttt{--enable-quic}. We have created our I-Frame element to embed a YouTube player inside the chrome browser by appending the YouTube video ID at the end of ``\url{https://www.youtube.com/embed}'', which is called as I-Frame SRC URL~\cite{YouTubeAPI}. We also set the auto-play parameter inside this I-Frame to play the video automatically once the player is loaded.

\begin{wraptable}{r}{0.5\textwidth}
\centering
\caption{Details of the Selected Videos}
\label{video-details}
\scriptsize
\begin{tabular}{|p{2cm}|p{3cm}|}
\hline
\textbf{Number of videos} & \textit{46} \\ \hline
\textbf{Video duration} & \textit{40 minutes - 1 hour 40 minutes} \\ \hline
\textbf{Genres of videos} & \textit{News, Entertainment shows, Education, Indian talk shows, Comedy, Stanford online lectures, British TV series} \\ \hline
\textbf{Minimum video quality achieved} & \textit{144p} \\ \hline
\textbf{Maximum video quality achieved} & \textit{720p} \\ \hline
\end{tabular}
\end{wraptable}

To ensure that video is not played from the cache, we created a new temporary user for opening a new Chrome window after clearing the cache; therefore, a Google Chrome window opens and starts running a particular video directly from YouTube. Note that each video was played twice in the Chrome browser, once with QUIC protocol enabled by setting the \texttt{--enable-quic} flag, and once with HTTP/1.1 (TCP) as the browser protocol. We verified this using the packet tracing tool, i.e., Wireshark, whether indeed TCP and QUIC packets were getting transferred. Initially, we observed that our institute's firewall blocks all QUIC traffic; hence even though we specified to run over QUIC, it ran over TCP protocol completely. We understood that this is due to the fallback option supported by QUIC to sustain across middleboxes. Next, we made an exception in the firewall rules and allowed QUIC traffic. We also checked that QUIC packets' sizes are smaller than the path's MTU. Additionally, we used an ad-block extension to remove advertisements during the video playback.

\subsection{Network Setup} 
Network bandwidth directly impacts YouTube ABR decisions~\cite{mondal2017candid}. Further, our objective is to analyze and compare the performance of a YouTube streaming session over a QUIC-enabled browser (say, $S_Q$) and a legacy HTTP/1.1 browser (say, $S_T$). For this purpose, we need to ensure that the network bandwidth patterns remain similar over the two streaming sessions $S_Q$ and $S_T$. While it is not possible to ensure replay bandwidth patterns over an in-the-wild setup, a large number of recent studies~\cite{mao2017neural,dong2018pcc,nejati2016depth,zheng2021xlink,arun2021toward,cangialosi2021site} have relied on benchmark network emulator frameworks like \textit{Mahimahi}~\cite{mahimahi} to analyze the protocol performance in a realistic setup. Accordingly, to emulate a specific network bandwidth over the test setup, we use \textit{Mahimahi} Link Shell (\texttt{mm-link}) emulation, which emulates network links using user-specified packet delivery trace files. Mahimahi maintains two queues -- one for the uplink traffic and the second for the downlink traffic. Whenever packets arrive from the Mahimahi's \texttt{mm-link} or Internet, it is placed directly into one of two packet queues depending upon whether it is uplink or downlink. Then it releases the packets based on input packet-delivery trace. So, each line in a trace file represents the time at which the packet of size MTU can be delivered. Also, \texttt{mm-link} wraps to the beginning of the input trace file on reaching its end. 


We have emulated various bandwidth patterns as shown in Table~\ref{bandwidth-pattern}. We start with static bandwidth to understand the minimum and the maximum bandwidth at which the minimum and the maximum video quality can be attained. At $<128$ Kbps, the lowest video quality $144p$ was observed throughout the video; similarly, at $>1500$ Kbps, the quality level $720p$ was observed. Accordingly, we finally used five different bandwidth patterns to generate the traffic traces, as listed in Table~\ref{bandwidth-pattern}. Each of the patterns initiates a loop from a starting bandwidth, makes a jump to the next bandwidth label, maintains them for a duration, and makes a jump again. This goes up to the last bandwidth as mentioned in Table~\ref{bandwidth-pattern}, and then the patterns repeat with the last bandwidth as the starting bandwidth, the starting bandwidth as the last bandwidth, and a negative value of the jump.

\begin{table}[t]
\scriptsize
\caption{Bandwidth Patterns}
\resizebox{12cm}{!}{
\begin{tabular}{|l|c|c|c|c|}
\hline
\textbf{Bandwidth Pattern} & \textbf{Starting} & \textbf{Last} & \textbf{Jump} & \textbf{Jump} \\ 
& \textbf{Bandwidth} & \textbf{Bandwidth} & \textbf{(Kbps)} & \textbf{Duration} \\ \hline
\textit{\textbf{Increasing (Incr)}} & 128Kbps & 2048Kbps & +64 & 60 sec \\ \hline
\textit{\textbf{Decreasing (Decr)}} & 2048Kbps & 128Kbps & -64 & 60 sec \\ \hline
\textit{\textbf{Dynamic High (DH)}} & 1152Kbps & 896Kbps & -256 & 240 sec \\ \hline
\textit{\textbf{Dynamic Low (DL)}} & 640Kbps & 128Kbps & -256 & 240 sec \\ \hline
\textit{\textbf{Dynamic Very Low (DVL)}} & 64Kbps & 256Kbps & +64 & 60 sec \\ \hline
\end{tabular}
}
\label{bandwidth-pattern}
\end{table}

Finally, from the individual bandwidth pattern, we generate emulated traces as follows, which are fed to the Mahimahi traffic shaper. We consider each packet size equal to the path MTU, i.e., $1500$ bytes. From this, we compute the send timestamp for each packet. Say, at time $T_i$, a packet has been transmitted; then the transmission time for the next packet $T_{i+1}$ is computed from the bandwidth at that time and the packet size (which equals to path MTU). For example, if the bandwidth is $640000$ bps, then $T_{i+1} = T_i + ((1500 * 8) / 640000)$. This trace is fed to the Mahimahi traffic shaper. To ensure that the end-to-end Internet bandwidth follows this emulated bandwidth pattern and the backbone bandwidth does not impact the YouTube performance, we performed a sanity check using a parallel streaming without applying any traffic shaping, as shown in Fig.~\ref{exp-setup}. We kept on observing that the same video, when streamed over the same network gateway but from a different machine without running the traffic shaper, always attains the maximum supported quality level.

\begin{table}[!t]
\caption{Instance of video information file of video id: -SI0HKTfHN4}
\resizebox{\textwidth}{!}{%
\begin{tabular}{|l|l|l|l|l|l|l|l|l|l|l|l|l|l|l|l|}
\hline
\textbf{Itag}               & 136     & 247     & 398     & 135    & 244    & 397    & 134    & 243    & 396    & 133    & 242    & 395    & 160    & 278    & 394   \\ \hline
\textbf{Bitrate (bits/sec)} & 2029085 & 1908018 & 1497188 & 930845 & 849429 & 737405 & 654628 & 492803 & 399554 & 301679 & 244471 & 213957 & 121862 & 112110 & 94436 \\ \hline
\textbf{Quality Label}      & 720p    & 720p    & 720p    & 480p   & 480p   & 480p   & 360p   & 360p   & 360p   & 240p   & 240p   & 240p   & 144p   & 144p   & 144p  \\ \hline
\end{tabular}%
}
\label{video-info}
\end{table}

\subsection{Log Collection Setup}  
We collect logs at two layers: the application layer and the network layer. For the application layer, we download all the video-related information from the requests made by YouTube client, and the responses received. Since the connections were encrypted, we needed to log these requests/responses within the browser; we created our own application log extension. Inside the application log extension, we have used the \texttt{console.log} API for the logs. We defined four events when we collected the logs -- \texttt{BEFORE\_REQUEST} (the YouTube client prepares an HTTP Request), \texttt{SEND\_HEADERS} (YouTube client sends the HTTP Request to the YouTube server, which contains various DASH parameters), \texttt{RESPONSE\_STARTED} (YouTube client starts receiving the response from the YouTube server) and \texttt{COMPLETE} (YouTube client completes receiving the response, including the video segment data, from the server). These events log all the HTTP Request messages that have been sent from the client to the server. 

For this events log, we observe two different types of requests -- a \texttt{videoplayback} request (request URL is like \url{https://r6---sn-o3o-qxal.googlevideo.com/videoplayback}) which is a HTTP GET Request, and a \texttt{qoe} request (Request URL is like \url{https://www.youtube.com/api/stats/qoe?...}) which is a HTTP POST Request. The \texttt{videoplayback} request contains the details of the requested video segment, such as \textit{the request timestamp} of the requested segment, \textit{total bytes} in the segment, byte \textit{range} of the segment data to be downloaded, \textit{itag} value (tells the requested video and audio quality), \textit{rbuf} (receiver buffer) in second, \textit{rbuf} in bytes, \textit{clen} (the maximum possible length of the downloaded segment), \textit{dur} (duration of the downloaded segment), the \textit{download speed}, and about the protocol it has used (either QUIC or TCP). Once the YouTube server receives this request, it sends the corresponding HTTP Response and the segment data as requested. 

On the contrary, the YouTube client periodically sends the \texttt{qoe} request to the server, on triggering of the event \texttt{streamingstat} that is triggered after a predefined number of frames are rendered over the client. This is a POST request that embeds the statistics about the video streaming, i.e., what amount of data has been rendered or played over the YouTube client. The \texttt{qoe} request embeds the following information: \textit{request timestamp}, \textit{itag} of the segment played, \textit{health of buffer} that tells at real-time $t$ for how much duration the video has been buffered, and \textit{cmt} that tells at real-time $t$ for how much duration the video has been played. We combine these two statistics to obtain the details of a video frame, i.e., when a frame has been downloaded, at what quality it has been downloaded, and when the frame has been played finally. 

It can be noted that YouTube \textit{itag} value encodes the bitrate in terms of quality labels, such as $144p$, $240p$, $360p$, $480p$, $720p$, etc. However, YouTube stores a mapping among $itag$, quality labels, and the corresponding bitrate in terms of a \texttt{video-info} file, which can be obtained through YouTube developers API. Table~\ref{video-info} shows the mapping for such a sample video. Indeed, for our analysis, we select the videos that contain such mapping in their \texttt{video-info} file. Interestingly, this covers both the \textit{constant bitrate} (CBR) and \textit{variable bitrate} (VBR) encoded videos, as the quality label to bitrate mapping can directly be inferred from a given \textit{itag} value. For example, Table~\ref{video-info} shows three different bitrates for each quality label, although the corresponding \textit{itag}s are different. In addition to the application logs, we also collected packet traces to get the network-level exchanges. We extract the information like the total amount of TCP bytes and the total amount of QUIC bytes transferred during a video streaming session.

\begin{figure}[t]
    \centering
    \includegraphics[width=0.7\textwidth]{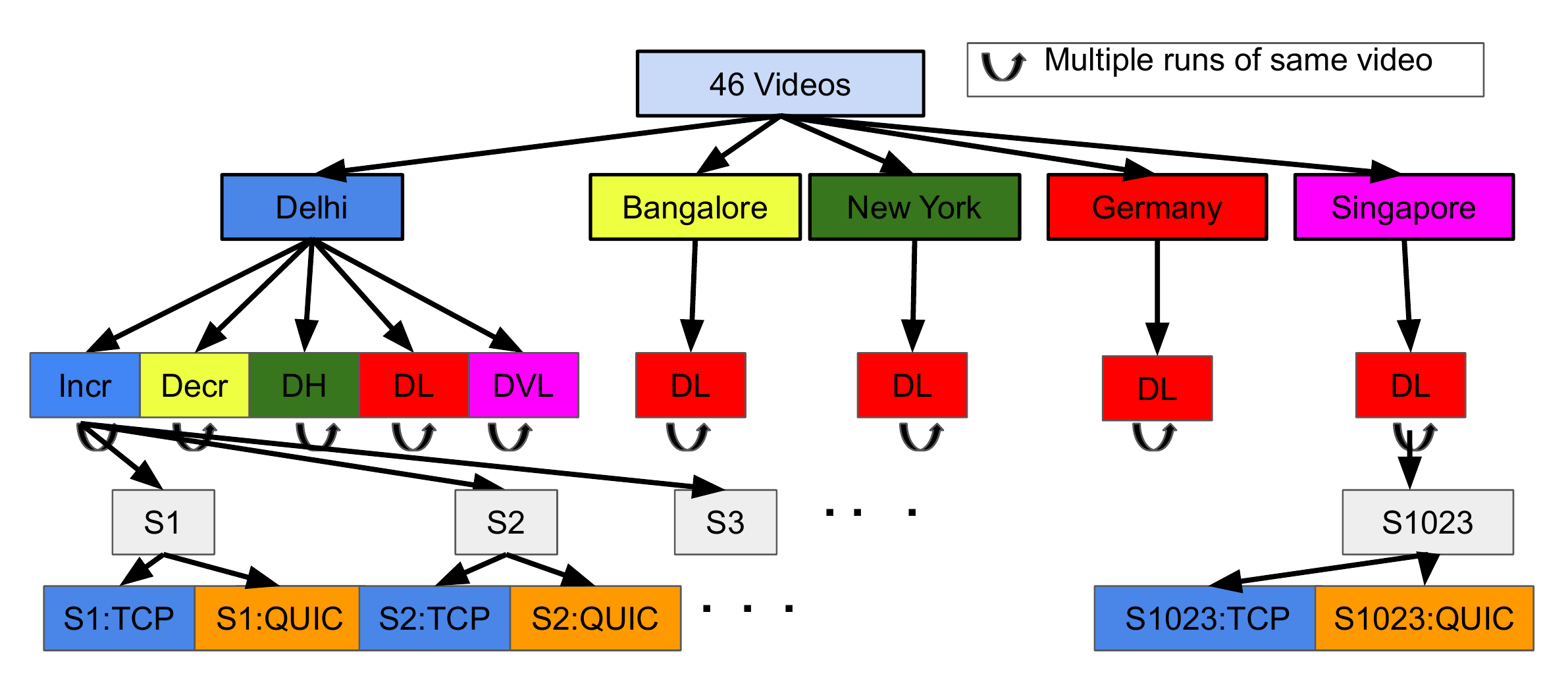}
    \caption{Dataset Organization}
    \label{dataset-hierarchy}
    \end{figure}

\subsection{Dataset Organization}
\begin{wraptable}{r}{0.6\textwidth}
 \centering
 \scriptsize
 \caption{Dataset details}
 \begin{tabular}{|l|l|}
 \hline
\textbf{Total streaming sessions} & $2046$ \\ \hline
 \textbf{Total duration} & $9555257$ seconds \\ \hline
 \textbf{Total TCP duration} & $4725312$ seconds \\ \hline
 \textbf{Total QUIC duration} & $4829945$ seconds \\ \hline
 \textbf{Incr duration} & $51676$ sec \\ \hline
 \textbf{Decr duration} & $47982$ sec \\ \hline
 \textbf{DH duration} & $227052$ sec \\ \hline
 \textbf{DL duration} & $6161572$ sec \\ \hline
 \textbf{DL Delhi} & $3150210$ sec \\ \hline
 \textbf{DL Bangalore} & $726267$ sec \\ \hline
 \textbf{DL New York} & $861046$ sec \\ \hline
 \textbf{DL Germany} & $893534$ sec \\ \hline
 \textbf{DL Singapore} & $530515$ sec \\ \hline
 \textbf{DVL duration} & $3066975$ sec \\ \hline
 \end{tabular}
 \label{dataset}
 \end{wraptable}Fig.~\ref{dataset-hierarchy} represents our entire dataset in a hierarchical organization. We have a list of $46$ videos of different genres. We run them across $5$ different geographical locations, namely Delhi, Bangalore, New York, Germany, and Singapore using Digital Ocean Machines~\cite{Do}.
 In Delhi, we ran videos for $5$ different bandwidth patterns (shown in Table.~\ref{bandwidth-pattern}). Since we are interested in the adaptation behavior of QUIC in the face of connection failures, we found out low bandwidth becomes a major concern there. Hence, we ran the video only for low bandwidth at other locations, i.e., DL. Again at each bandwidth, we ran a video multiple times. In total, we obtain datasets of the total of $1023$ steaming session pairs that we name as \textit{S1, S2, ...., S1023}. Each session pair contains one session over a QUIC-enabled browser and another for the legacy HTTP/2 browser with pure TCP connections. We summarize the dataset collected in Table.~\ref{dataset}. The total duration of video logs is $9555257$ seconds, which corresponds to about $2655$ hours of video streaming sessions. Out of these total data, for \textit{Incr} bandwidth pattern the dataset is of $51676$ sec of streaming duration and for \textit{decr} bandwidth pattern the dataset is of $47982$ sec of streaming duration, for \textit{DH} pattern, it is of $227052$ sec of streaming duration. Similarly, for \textit{DL} bandwidth pattern, the total duration is of streaming is $6161572$ sec; whereas, for \textit{DVL} bandwidth pattern, it is $3066975$ sec.


\subsection{Metrics Used for Analysis}
\begin{wrapfigure}{l}{0.3\textwidth}
\includegraphics[width=0.3\textwidth]{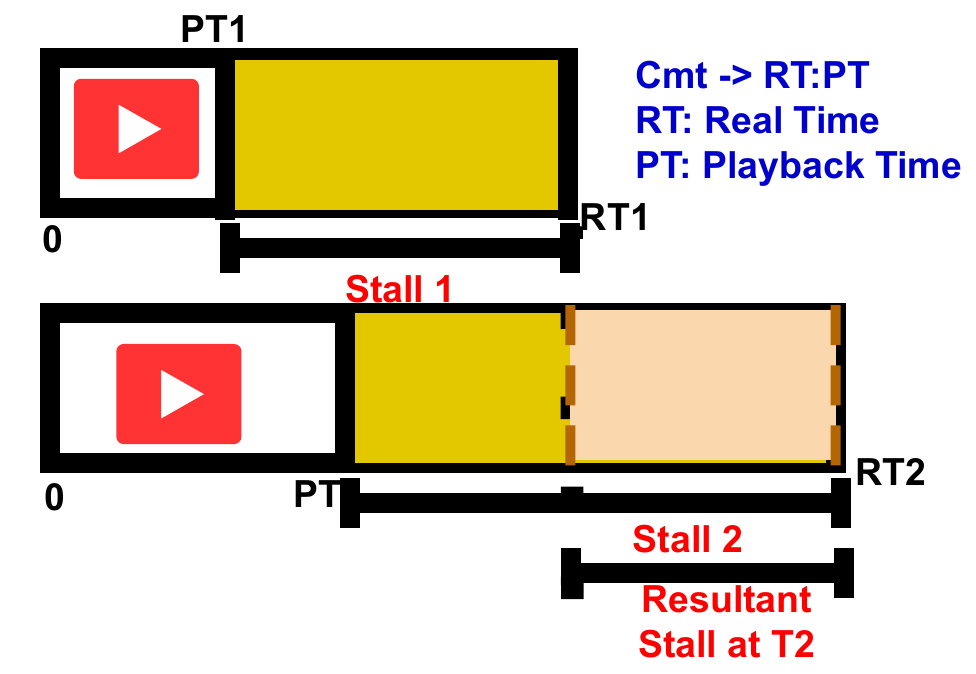}
\caption{Stall computation}
\label{stall-comp}
\end{wrapfigure}
In this paper, we adopted the formula-based QoE metric as used by Penseive~\cite{mao2017neural}. The QoE computation includes the average bitrate, average bitrate variation, and average stalling that the user experience during the video playback. We compute these parameters using the \textit{streaming-stat} information. Note that the HTTP response contains the information corresponding to a requested segment. To compute bitrate, we extract the \textit{itag} value inside the streaming-stat and then map it to the video information file as obtained from YouTube. We compute the bitrate variation by finding the difference between previous bitrate and the current bitrate. To compute stall, we take use of the \textit{cmt} parameter of the streaming-stat. We subtract the playback timestamp from the real timestamp (Fig.~\ref{stall-comp}), which gives the cumulative stall at any point of time. We then subtract the stall computed for the previous instance of the streaming-stat log from the current instance of the streaming-stat log to obtain the temporal distribution of the stall time. Given that the streaming-stat is asynchronous with the underlying events, we compute the average by taking all the samples from the beginning to the current instance of the streaming-stat. Consequently, we obtain a moving average of these three QoE parameters. Following Penseive~\cite{mao2017neural}, we compute the QoE as follows. 
\[\text{QoE} = \text{Average Bitrate} - \text{Average Bitrate Variation} - 4.3 \times \text{Average Stall} \]
We compute multiple QoE values (from multiple instances of streaming-stat) for each streaming session. We do not compute the average of these QoE values obtained in a streaming session, since it will not be able to  capture  instantaneous changes.

\section{QUIC and TCP: Dependability and Performance Implications}
\label{sec:analysis1}
In this section, we focus on understanding how a QUIC-supported browser utilizes QUIC and TCP simultaneously to fetch data from YouTube streaming server, and then analyze how it impacts the application's performance. It can be noted that we performed Welch's T-test on all the results and reported the corresponding $p$-values. The details follow.

\subsection{The Race between TCP and QUIC}
\begin{figure}[!t]
\subfloat[]{\includegraphics[width = 0.3\textwidth]{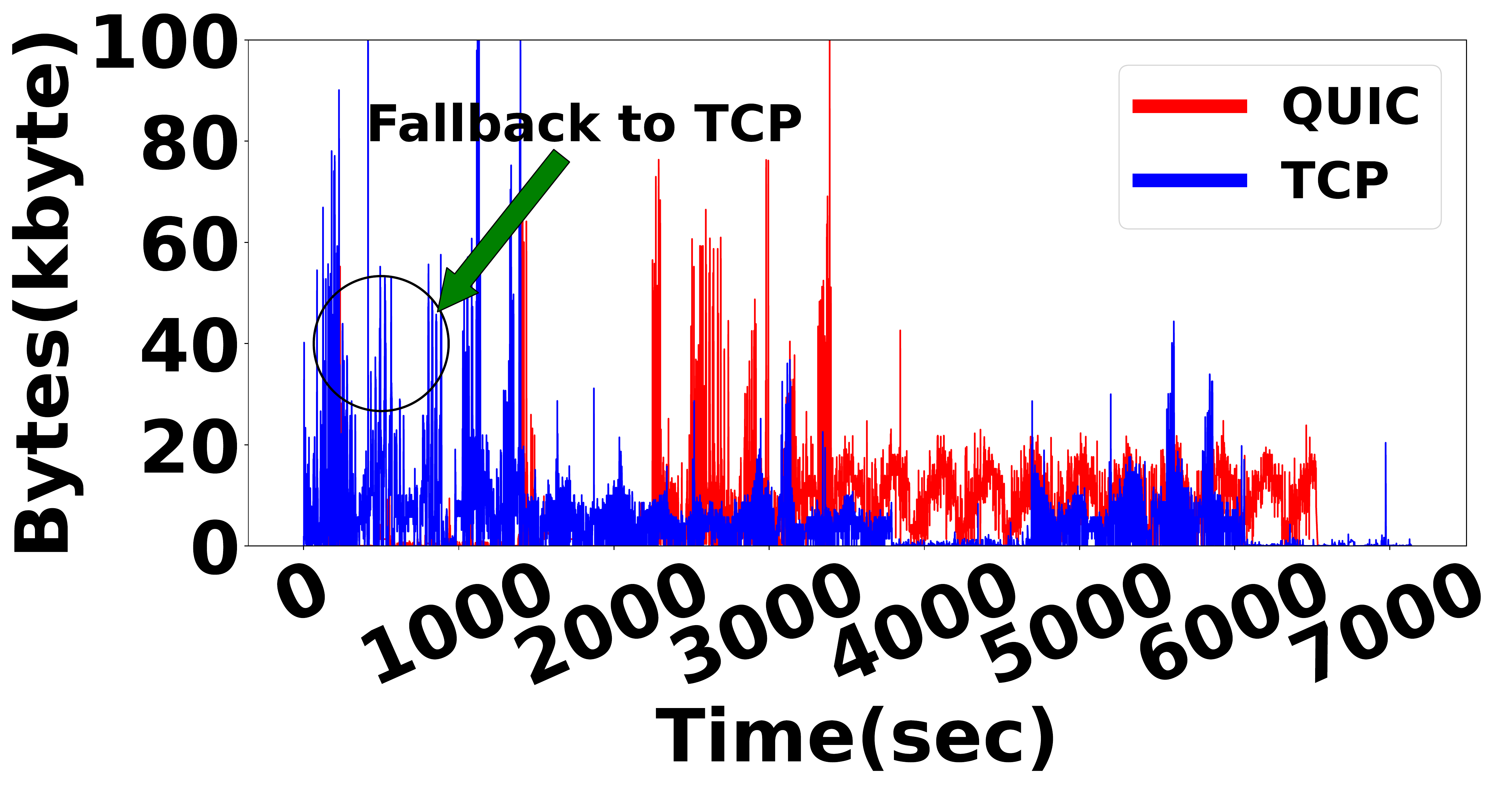}}
\subfloat[]{\includegraphics[width = 0.3\textwidth]{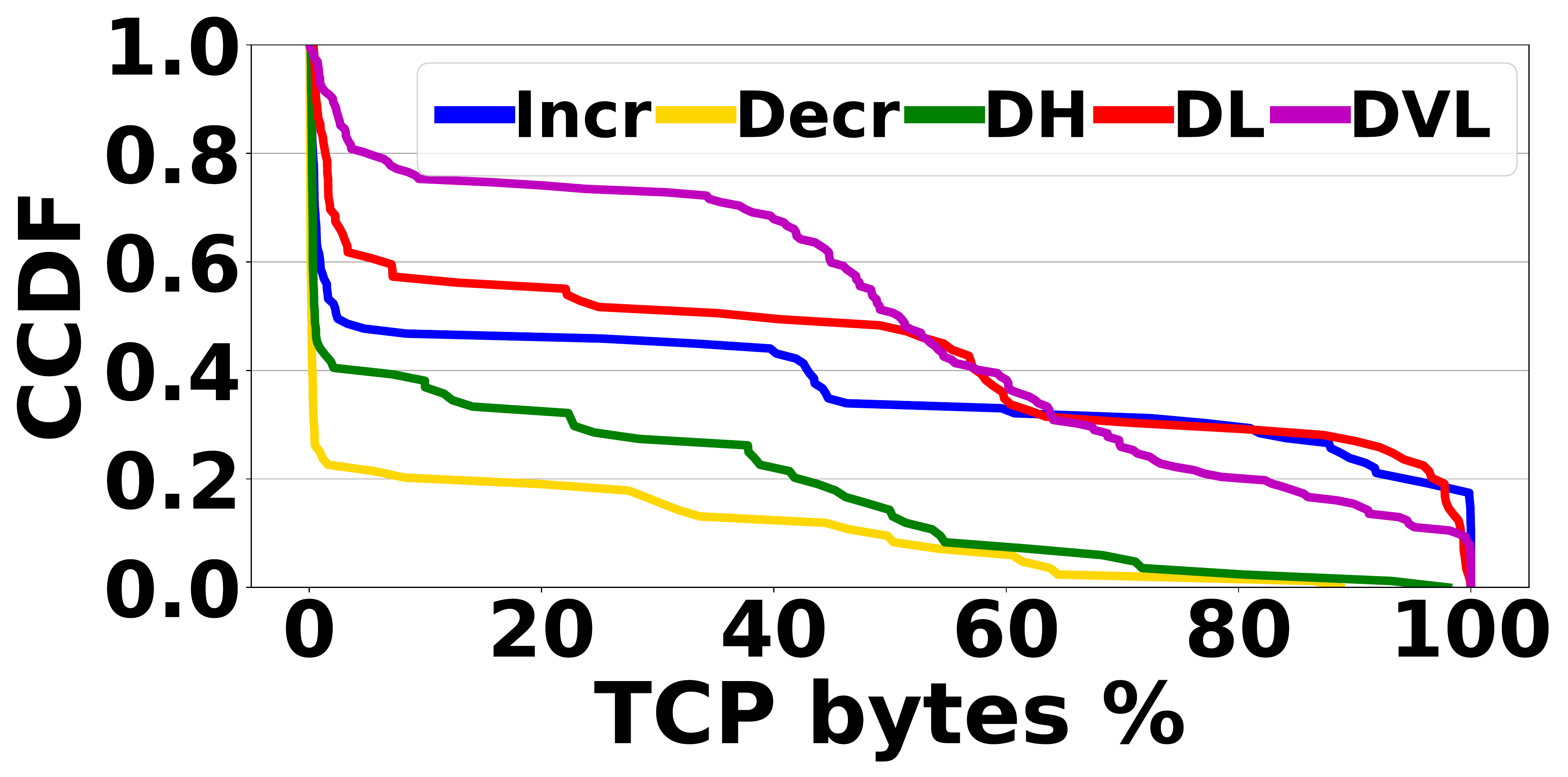}}
\subfloat[]{\includegraphics[width = 0.3\textwidth]{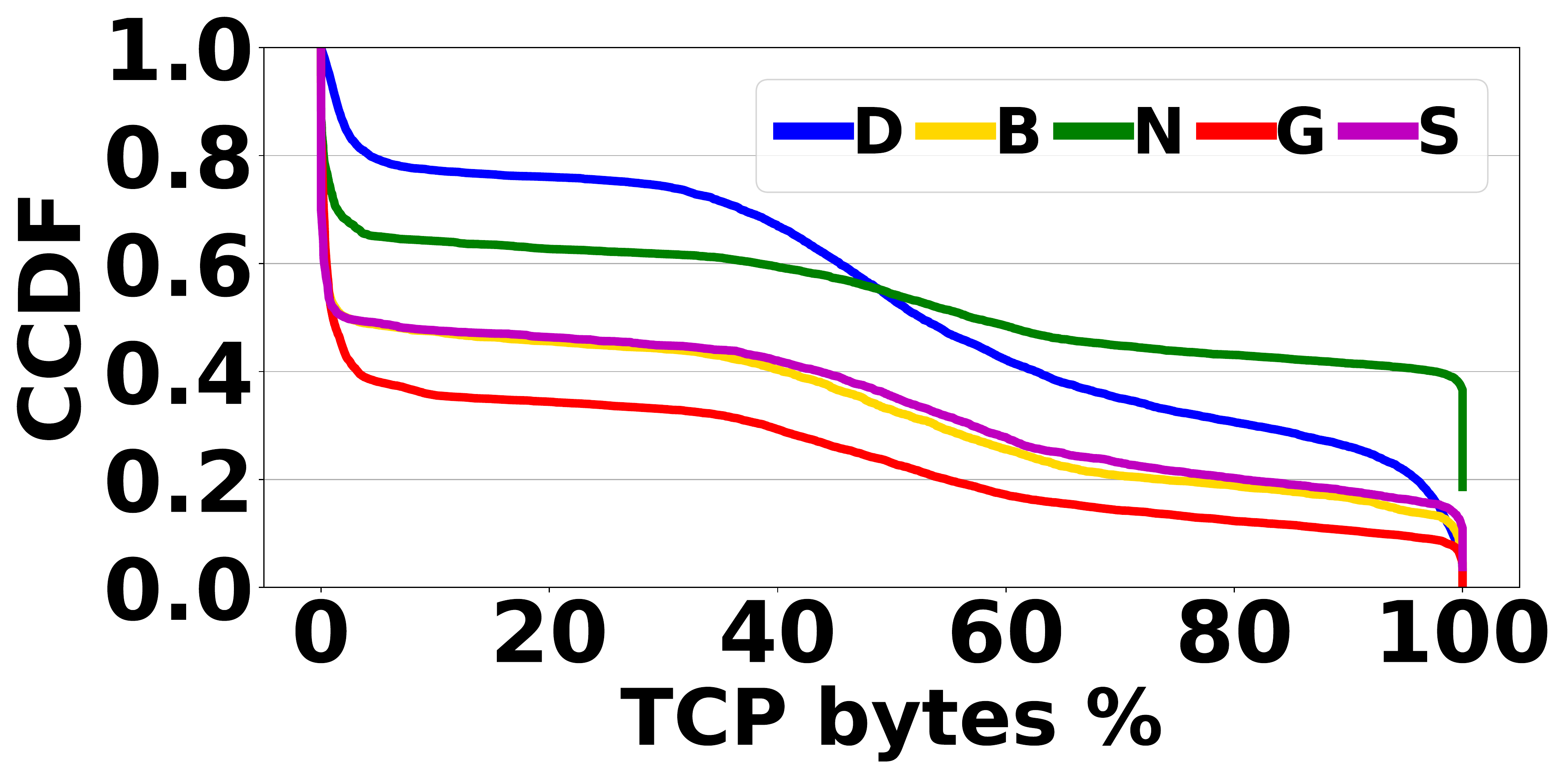}}
\caption{Fallback of QUIC to TCP (a) Number of bytes transferred over TCP and QUIC over a QUIC-enabled stream, (b) Percentage of fallback, i.e., percentage of bytes transferred over TCP in a QUIC-enabled stream for various bandwidth patterns, (c) Percentage of fallback, i.e., percentage of bytes transferred over TCP in a QUIC-enabled stream for various locations (D=Delhi, B=Bangalore, N=New Delhi, G=Germany, S=Singapore)}
\label{fallback-percent}
\end{figure}

As discussed earlier, to alleviate the impact of middleboxes, a QUIC-supported browser races a TCP connection with the QUIC connection. Fig.~\ref{fallback-percent}(a) shows the bytes transferred over both TCP and QUIC while a sample video was played by enabling QUIC. Note that the presence of TCP packets is inevitable; hence, it is expected that there will be the presence of TCP packets.  But, if QUIC wins the race, the presence of TCP packets is likely to be small as there are only TCP connection request packets then. We made an exception in the firewall rules of our Institute (discussed in Sec.~\ref{sec:expt-setup}). However, the figure shows that there is a heavy presence of TCP. This phenomenon indicates heavy fallback from QUIC to TCP. We have a few interesting observations here. (1) We notice that for every QUIC-enabled video session, the initial packet exchanges are over TCP only; QUIC does not even try to establish a connection initially for some duration. This duration was about $1$ sec for high bandwidth, and for low bandwidth, it was about $10-12$ sec (details in Appendix). (2) In the process of connection establishment, the client-server handshake packets go on multiple times before the connection is successfully established, where the client sends the handshake packet and an Encrypted Payload. For \textit{DH}, this process repeats $1-2$ times; for \textit{DL}, it repeats $2-3$ times; for \textit{DVL}, it repeats $3-4$ times before the successful connection establishment. We also noticed several extreme cases at DL and DVL, where the client cannot successfully send a handshake packet even after $3-4$ repetitions of this process. The client's IP layer then sends an ICMP error message to the server followed by TCP SYN packets (details in Appendix). At that point, QUIC is probably marked as broken for the server, and then there are no QUIC packets from the client to the server for around $600-1000$ sec. The fallback happens during this period when the application data is transferred over TCP. (3) Further, we observe that the client tries to establish a 0-RTT connection with the server. The server sends handshake packet to the client. But after that, client again sends initial request packet on the same port (details in Appendix). This defeats the purpose of 0-RTT connection support provided by QUIC.

%

Fig.~\ref{fallback-percent}(b) shows the complementary CDF (CCDF) of the percentage of TCP bytes experienced at various bandwidth patterns. We observe that for \textit{Incr} bandwidth, about $2$\% or more bytes are transferred over TCP for $50$\% times. For \textit{Decr} and \textit{DH}, there is almost negligible presence of TCP traffic for $78$\% times and $60$\% times, respectively. Indeed, it is expected that QUIC does not face many connection failures for high bandwidth. For \textit{DL} and \textit{DVL}, $40$\% or more bytes are transferred over TCP for $50$\% and $67$\% times, respectively. Hence, we observed that QUIC tends to fallback to TCP at a low-bandwidth network. This leads us to ask this question -- what if we had a pure TCP connection under such scenarios. \textit{Does the mixing of QUIC and TCP help an application perform better than pure TCP in terms of application QoE?} Further, \textit{does the QUIC-TCP-QUIC handover pose any additional overhead?} Henceforth, we use the terminologies \textit{QUIC-enabled stream} to denote a YouTube video streaming session with QUIC enabled in the browser, and \textit{TCP stream} to denote the same with QUIC disabled in the browser. It can be noted that a \textit{QUIC-enabled stream} may contain both the QUIC as well as TCP traffic depending on fallback instances, whereas a \textit{TCP stream} contains TCP traffic only.


\subsection{QUIC-enabled Stream vs TCP Stream -- How Do They Impact Application QoE?}

\begin{wrapfigure}{R}{0.5\textwidth}
    \subfloat[]{\includegraphics[width = 0.25\textwidth]{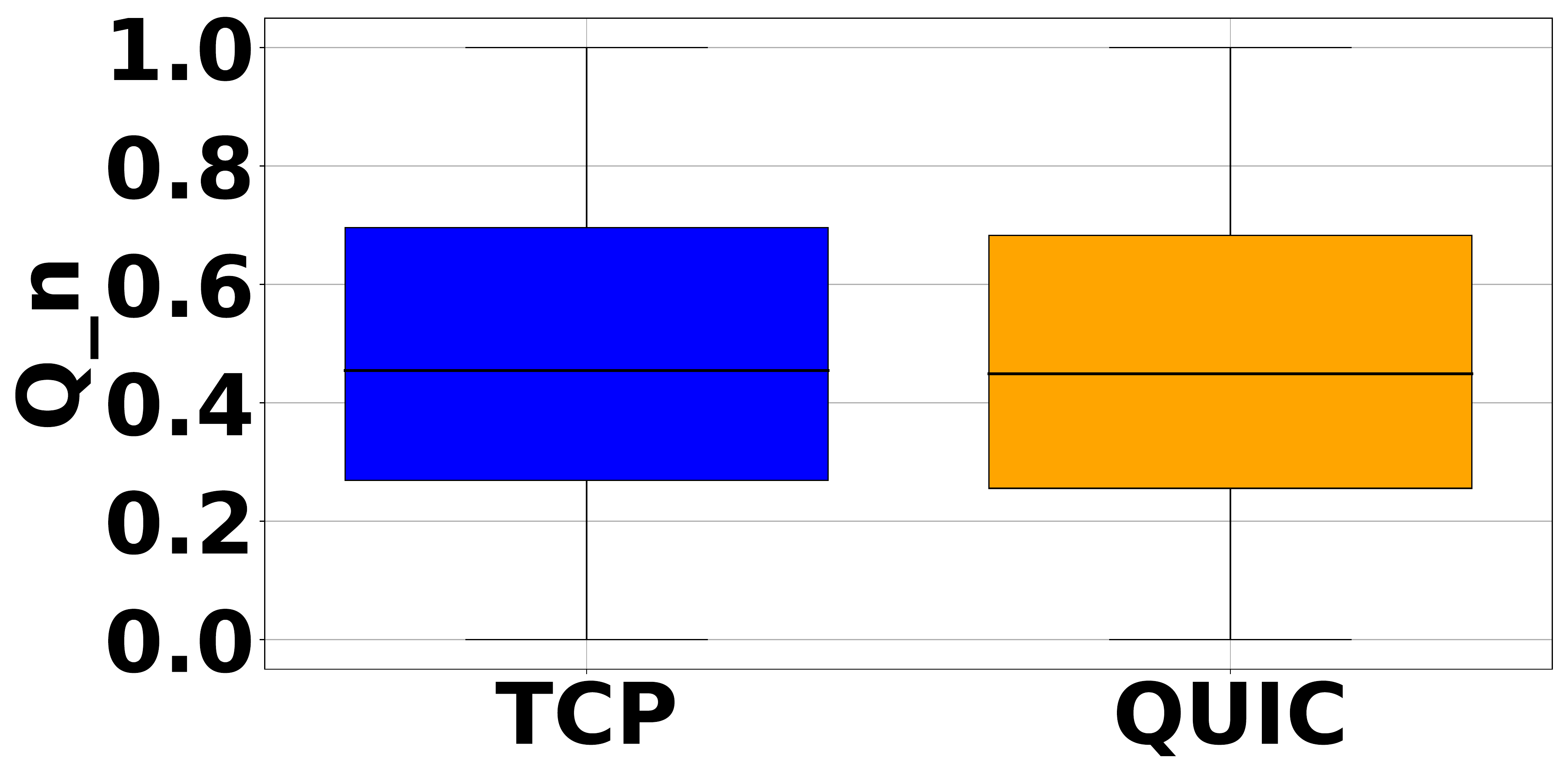}}
\subfloat[]{\includegraphics[width = 0.25\textwidth]{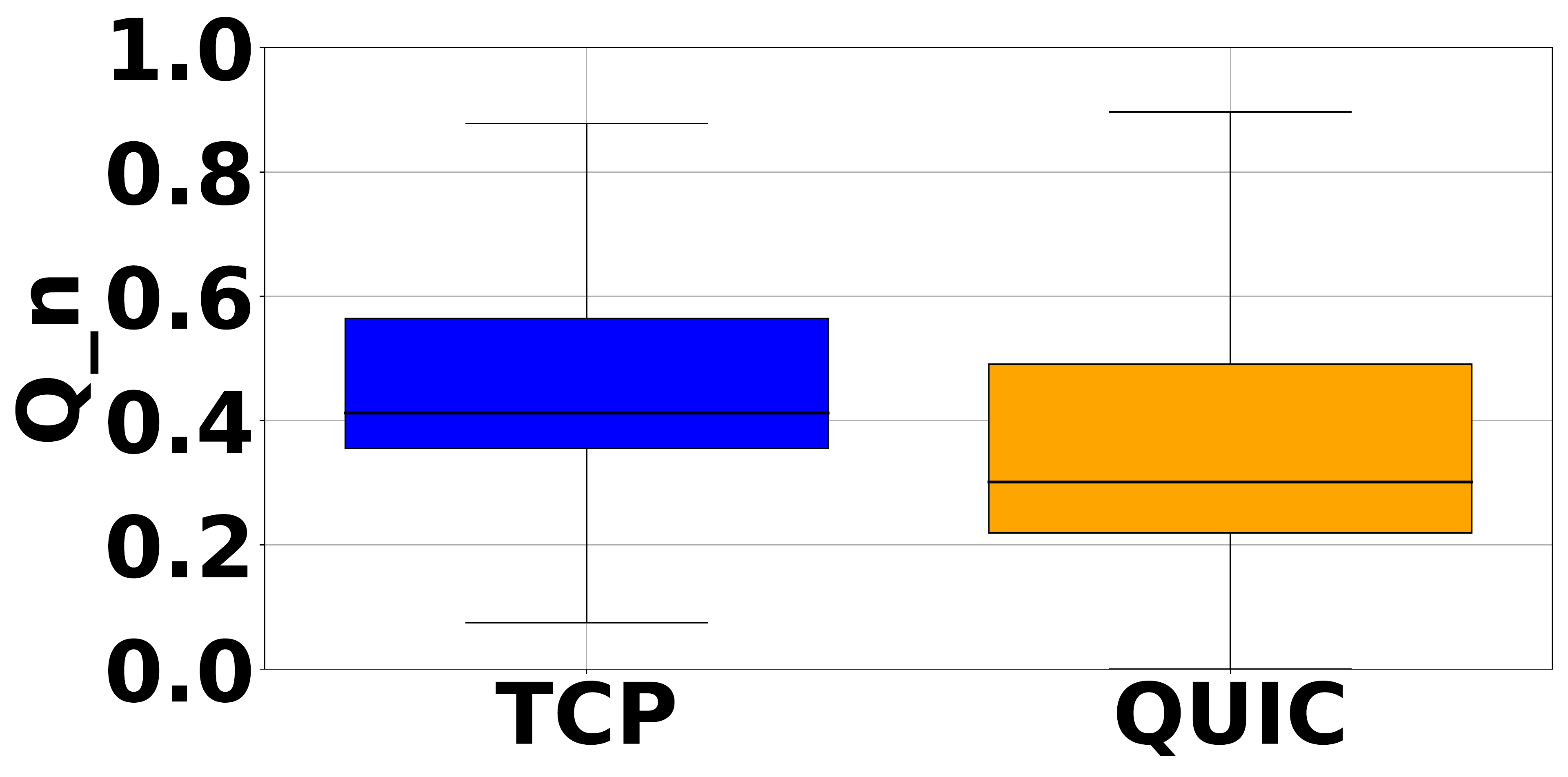}}
\caption{Normalized QoE ($Q\_n$) distribution for TCP and QUIC (a) across all $1023$ session pairs (b) for $2$ sessions (TCP and QUIC) for a sample session pair}
\label{QoE-distribution}
\end{wrapfigure}

To answer this question, we calculate the instantaneous QoE for each video. 
Note that an averaging approach of computing a single QoE value for each video is not a good approach as it fails to capture instantaneous changes (discussed in Sec~\ref{sec:expt-setup}). Hence, we compute multiple temporal QoE values for one video for both TCP and QUIC. This approach allowed us to capture instantaneous changes in the QoE parameters. Fig.~\ref{QoE-distribution}(a) shows the distribution of normalized QoE values for all $1023$ streaming pairs collected across $5$ locations and $5$ bandwidth patterns (See Fig.~\ref{dataset-hierarchy}).  Overall, we observe that there is a marginal difference in the application QoE when QUIC-enabled streams are used vs. when TCP streams are used for end-to-end transport. 

Although the average performance in long runs does not indicate any significant difference between QUIC-enabled streams versus TCP streams, we have a different observation if we analyze individual videos separately. We consider a sample case of a streaming session pair where the same traffic shaper is applied temporarily. Still, video streaming happens over TCP connections in one case, and in another case, we enable QUIC in the browser. From Fig.~\ref{QoE-distribution}(b), we observed differences in the perceived QoE between the two cases; we see that enabling QUIC in the browser degrades the video QoE performance (we observed statistical difference between the two distributions with a $p$-value of $<0.05$).  

\begin{figure}[!t]
\subfloat[]{\includegraphics[width = 0.25\textwidth]{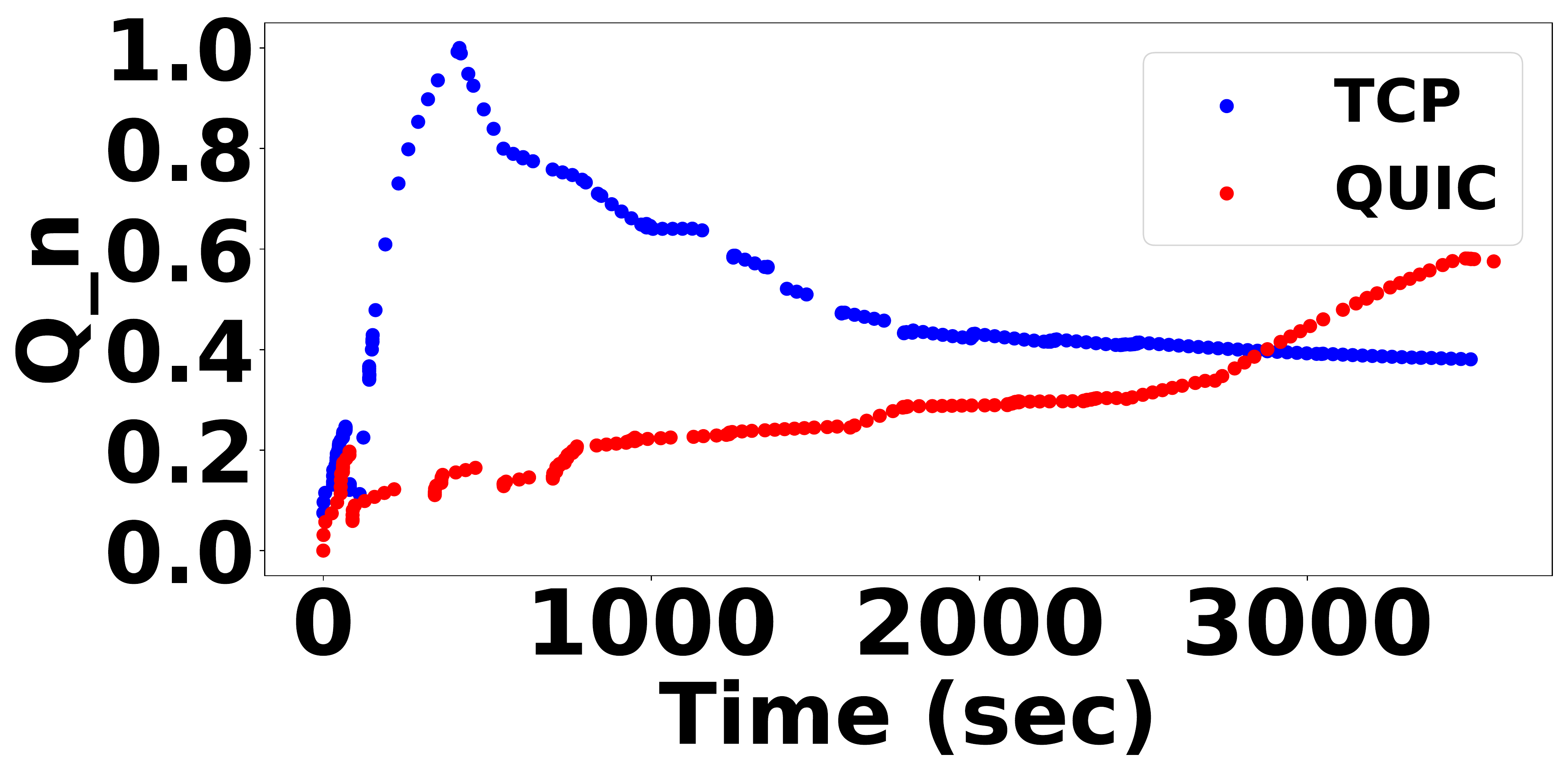}}
\subfloat[]{\includegraphics[width = 0.25\textwidth]{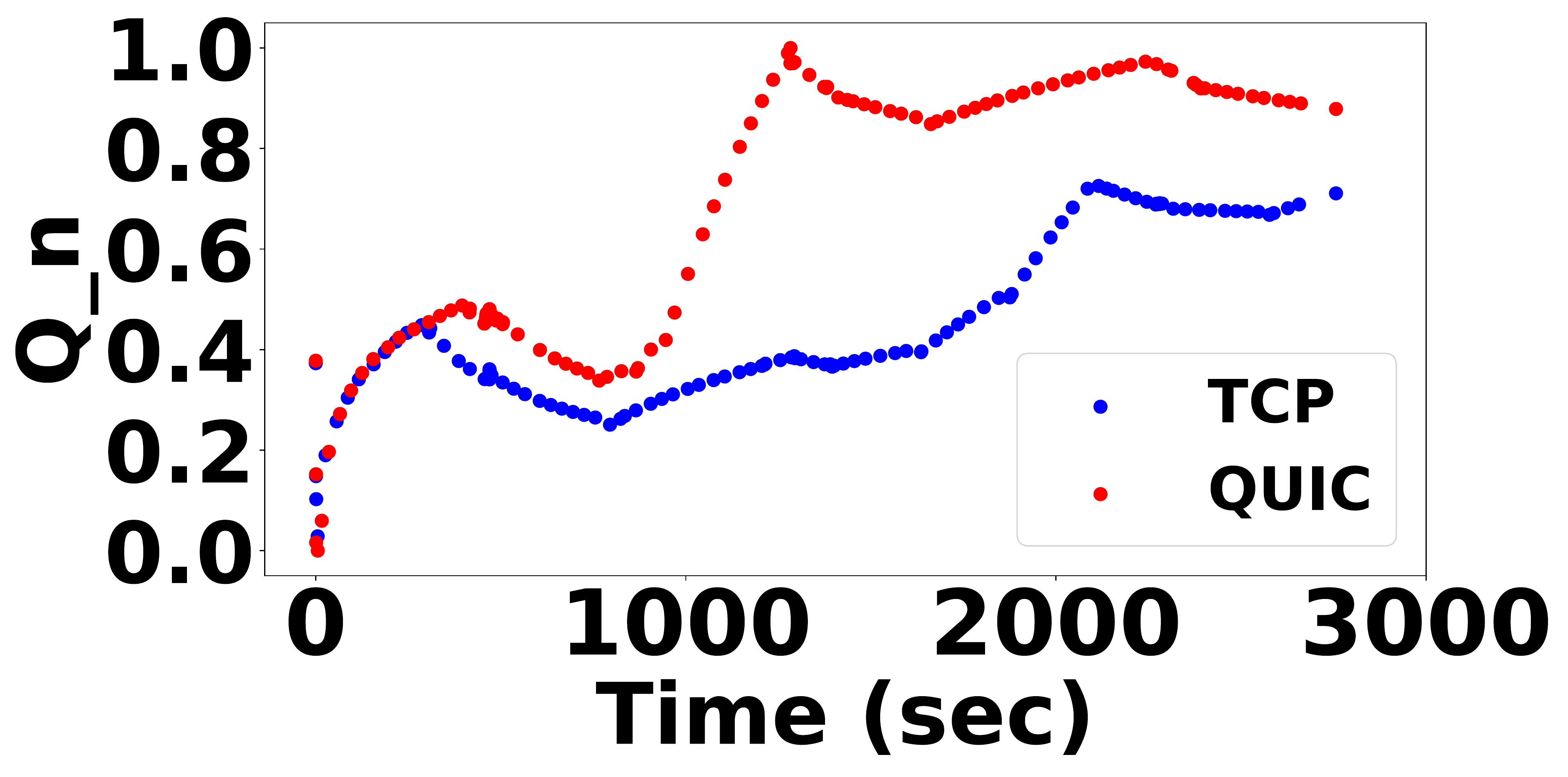}}
\subfloat[]{\includegraphics[width = 0.25\textwidth]{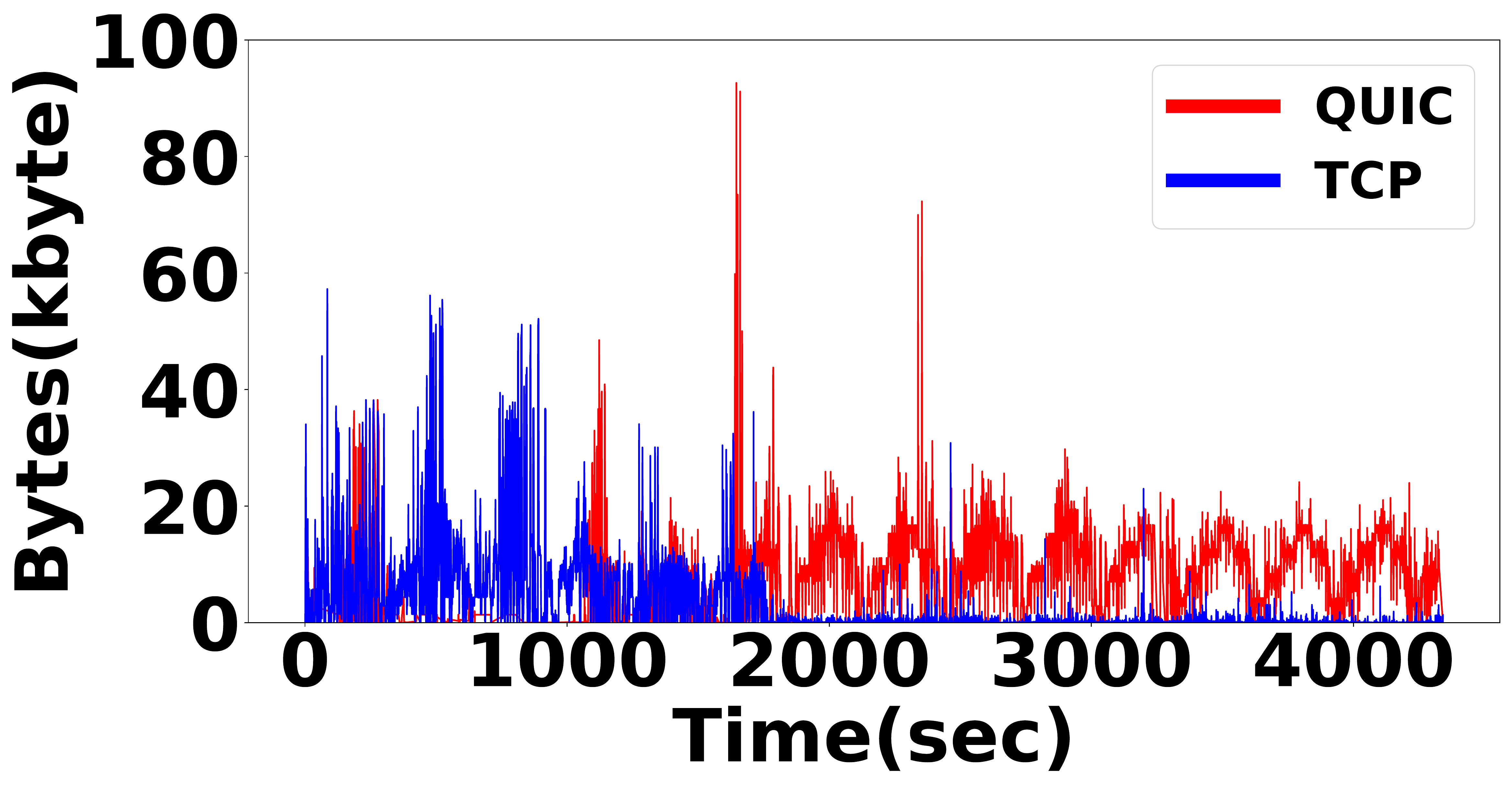}}
\subfloat[]{\includegraphics[width = 0.25\textwidth]{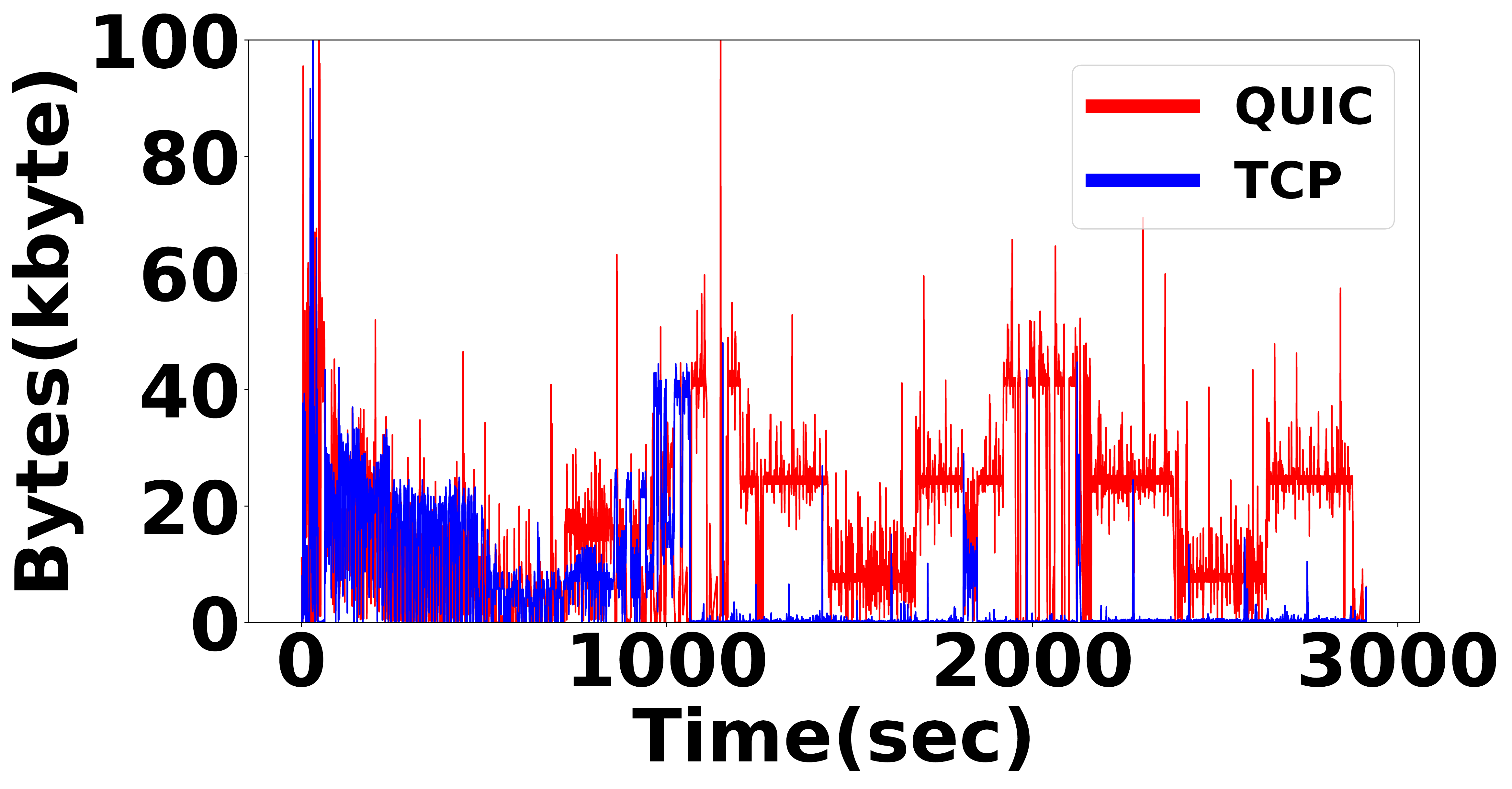}}
\caption{Normalized QoE ($QoE\_n$) values for a sample streaming session pair (a) where TCP stream performs better than QUIC-enabled stream, (b) where QUIC-enabled stream performs better than TCP stream; (c) Number of bytes transferred over TCP and QUIC for Case (a), (d) Number of bytes transferred over TCP and QUIC for Case (b)}
\label{qoe-scatter-plot}
\end{figure}

To analyze this further, we plot the perceived QoE values over the time for two pairs of sample video streaming sessions over two different cases. In the first case (Fig.~\ref{qoe-scatter-plot}(a)), the normalized QoE over the TCP stream is better (with $p < 0.05$) than that over QUIC-enabled stream under similar traffic shaping. In the second case (Fig.~\ref{qoe-scatter-plot}(b)), the normalized QoE over the QUIC-enabled stream is better (with $p < 0.05$) than that over the TCP stream. From Fig.~\ref{qoe-scatter-plot}(a), we observe that the temporal QoE has not been boosted up over the QUIC-enabled stream, whereas the TCP stream has been able to boost up the application performance (around time $500$ sec). To see the reason, we plotted temporal traffic patterns in Fig.~\ref{qoe-scatter-plot}(c) and (d) for the two cases when QUIC was enabled over the browser. We observe in Fig.~\ref{qoe-scatter-plot}(c) that when TCP stream performed better than QUIC-enabled one, the presence of TCP bytes is significant till up to $2000$ sec. There are quite a few peaks of more than $40$ KB of TCP. Whereas for the other case Fig.~\ref{qoe-scatter-plot}(d), the presence of TCP traffic is lesser than that of Fig.~\ref{qoe-scatter-plot}(c), the peaks are close to $20$ KB only. Moreover, it is significant only till about $500$ sec.

\subsection{How Often Does A QUIC-enabled Stream Perform Poorly?}
\begin{wrapfigure}{r}{0.5\textwidth}
    \subfloat[]{\includegraphics[width = 0.25\textwidth]{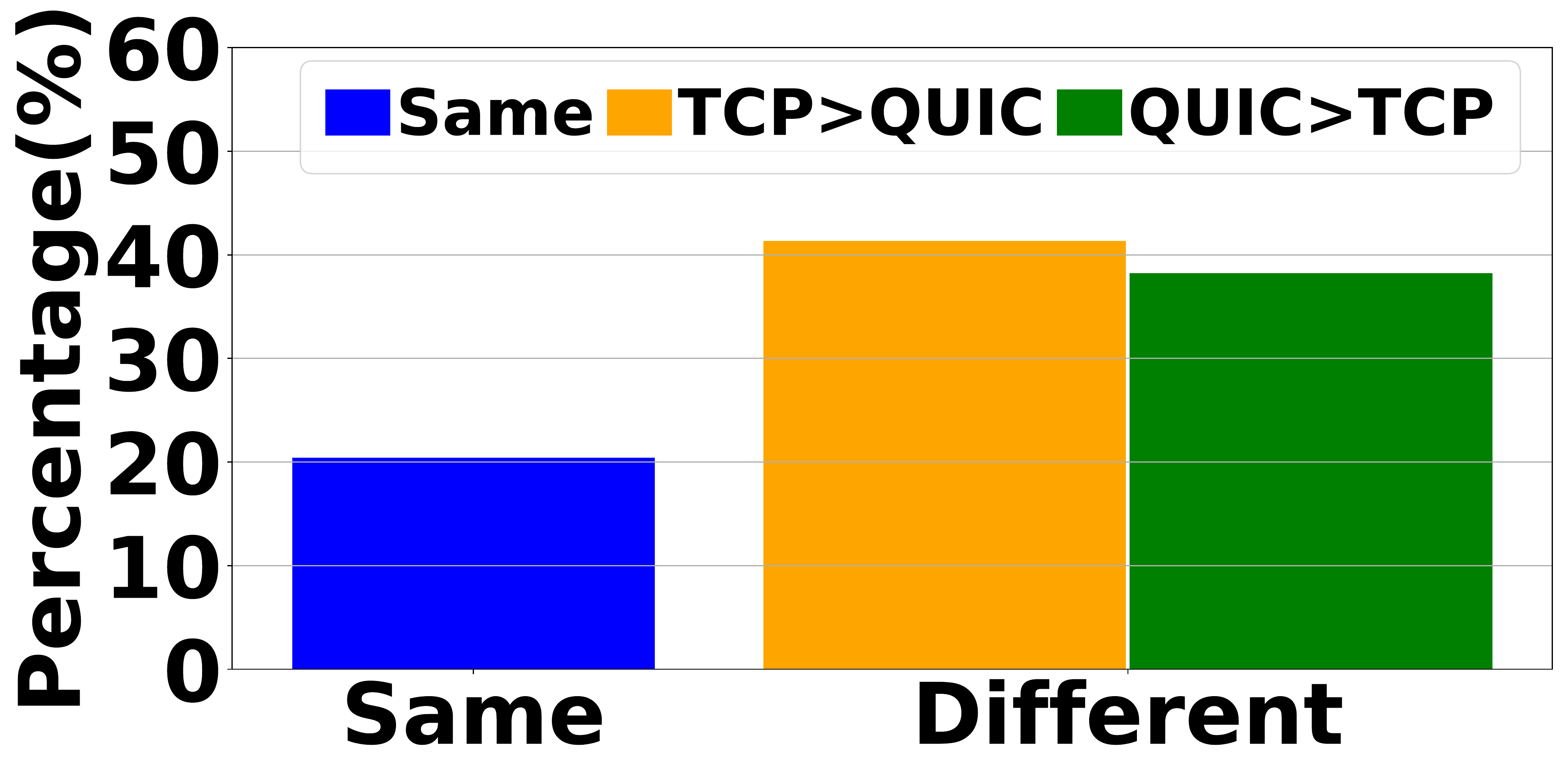}}
\subfloat[]{\includegraphics[width = 0.25\textwidth]{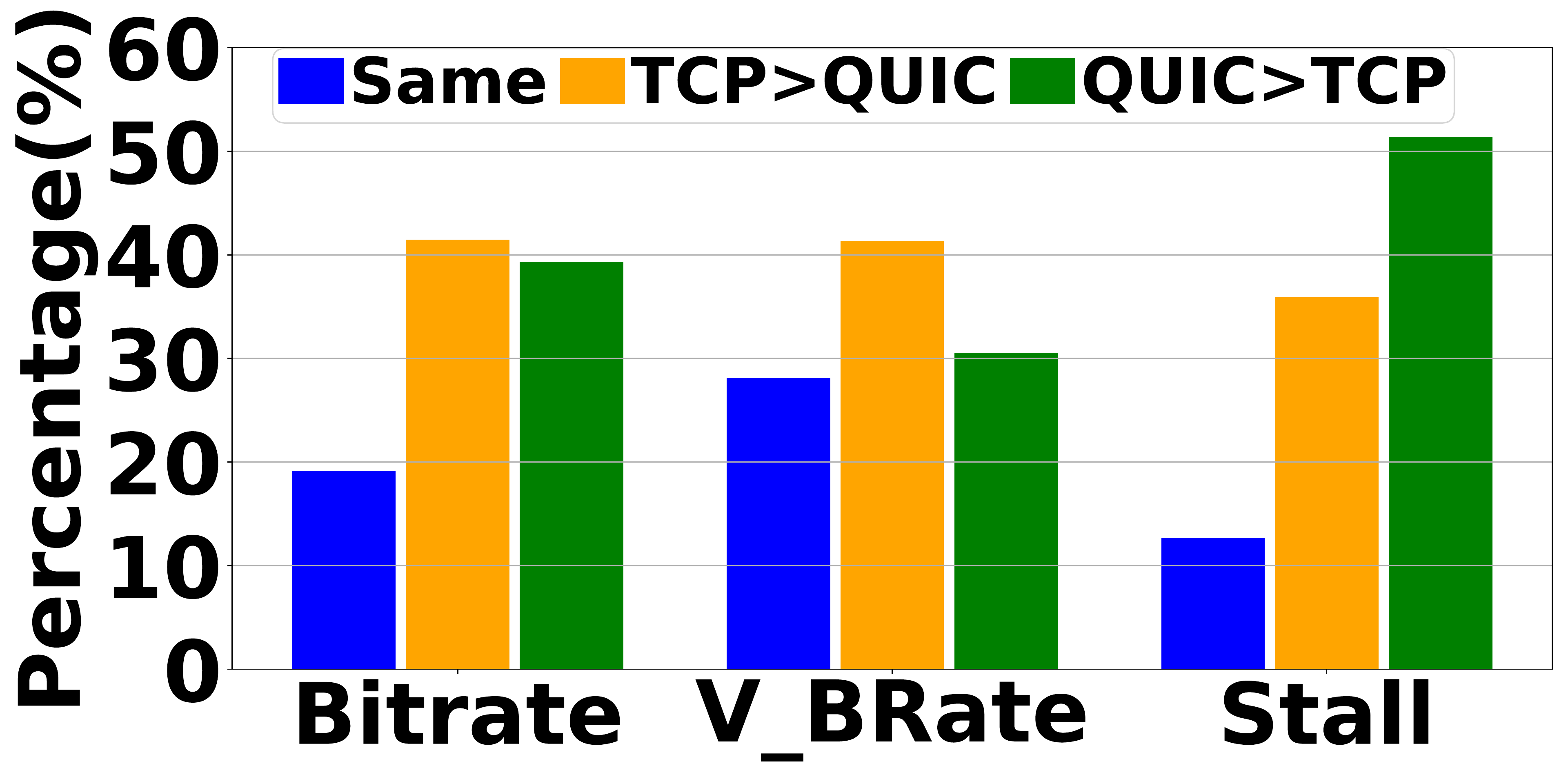}}
\caption{Hypothesis testing on (a) QoE (b) individual QoE parameters}
\label{hypothesis-result}
\end{wrapfigure}

We performed hypothesis testing over all the different streaming sessions to find out in what percentage of sessions a QUIC-enabled stream performs (a) better than, (b) statistically similar, and (c) worse than the corresponding TCP stream. We consider the case (b) as the null hypothesis and then perform a two-tail test~\cite{t-test} to check which protocol performs better when the alternative hypothesis is true. Fig.~\ref{hypothesis-result}(a) shows the result on all the video sessions for the overall normalized QoE. We observe that for $20.5$\% cases, the two streaming scenarios (QUIC-enabled and TCP) behave similarly, and for the remaining $79.5$\% of the cases, the two streaming scenarios behave differently. Further, For $41$\% cases, a TCP stream yields a better application QoE than 
the corresponding QUIC-enabled stream under a session pair. For $38.5$\% cases, a QUIC-enabled stream yields a better QoE than the corresponding TCP stream. To establish this fact further, we performed a hypothesis testing over all the QoE parameters over all the session pairs, as shown in Fig.~\ref{hypothesis-result}(b). We observe that for average playback bitrate and bitrate variation, TCP streams outperformed QUIC-enabled streams in $41$\% of the cases. However, the QUIC-enabled streams resulted in a higher stall compared to TCP streams for $51$\% of the instances, contradictory to previous studies~\cite{quicd,langley2017quic,chrome-blog} that claim that QUIC results in lesser stalls. The latest report~\cite{chrome-blog} says it results in $9$\% fewer stalls. This indeed motivates us to find out the root cause behind such poor performance of QUIC-enabled streams, particularly for the low-bandwidth scenarios.

From the above analysis, we observe that -- (a) depending on the network bandwidth, the instances of QUIC to TCP fallbacks are quite common when QUIC is enabled for end-to-end transport, and (b) although such fallback might not show significant impact on the video QoE for long run, they result in instantaneous drops in QoE and thus might affect the streaming performance for individual users. 
In other words, we observe that a protocol fallback over the QUIC-enabled stream impacts short-term application QoE. 
In the next section, we explore this in further details.

\section{Delving Into the Depths: Interplay of Different Parameters}
\label{sec:analysis2}
With various enhancement that QUIC enjoys, one would expect gain a better performance using QUIC compared to TCP. However, we observe that it did not happen for around $40\%$ of the cases. This led us to investigate further to see if this is due to any particular parameter or configurations that we have used or whether it is a general behavior of QUIC-enabled browsers. The detail follows.

\subsection{Do Individual QoE Parameters Show Similar Behaviour?}
\begin{figure}[!t]
\subfloat[]{\includegraphics[width = 0.25\textwidth]{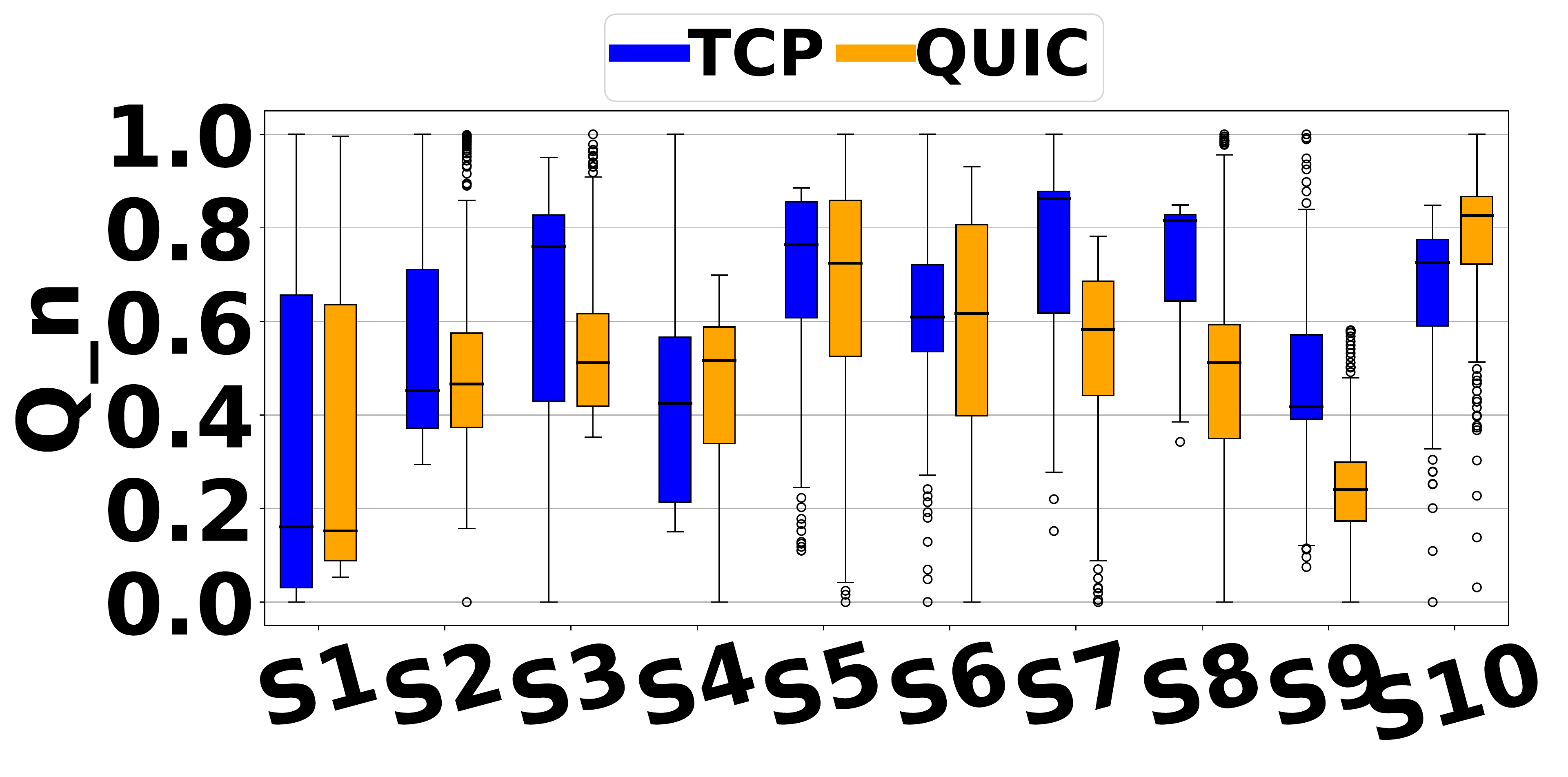}} 
\subfloat[]{\includegraphics[width = 0.25\textwidth]{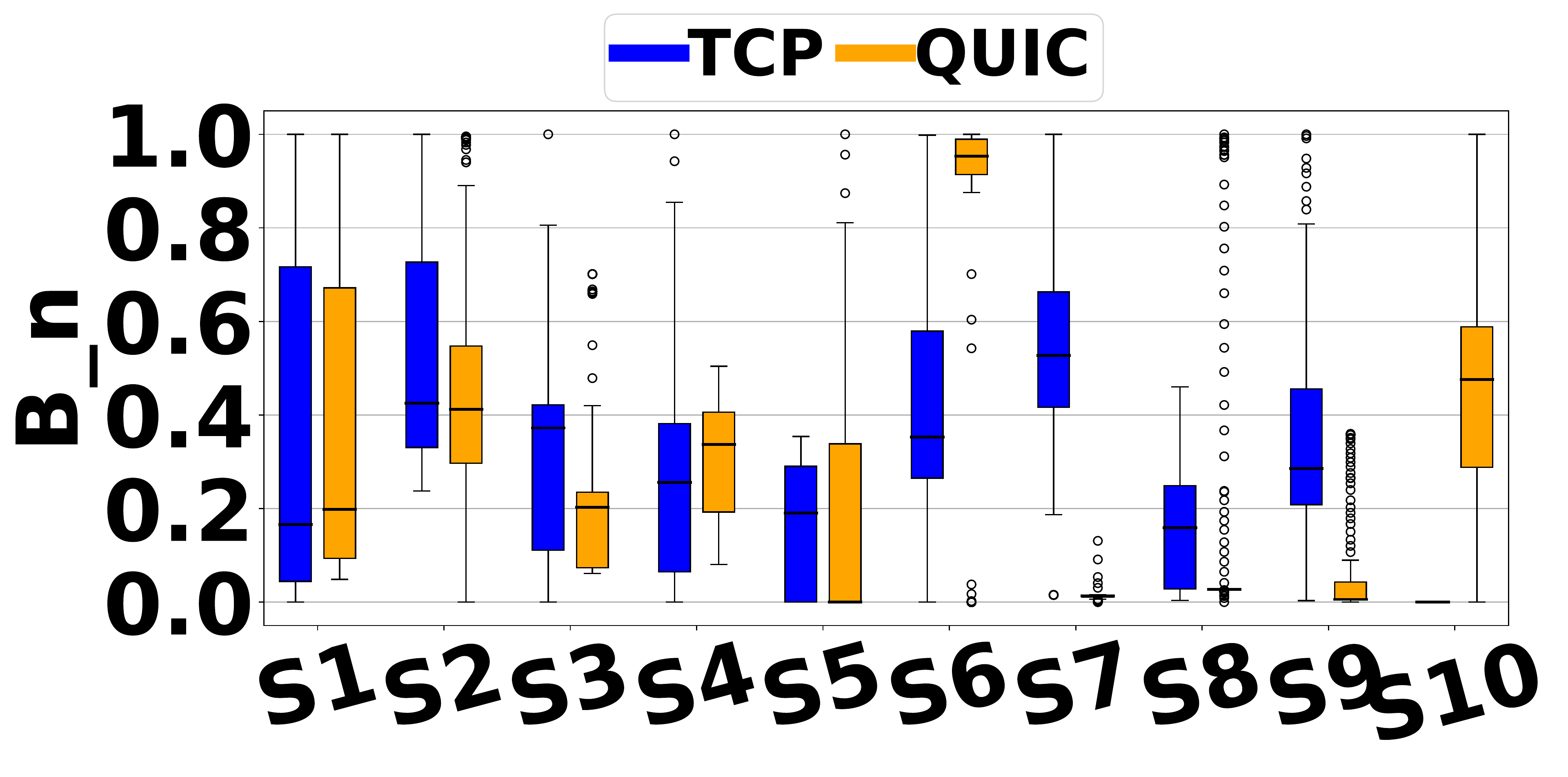}}
\subfloat[]{\includegraphics[width = 0.25\textwidth]{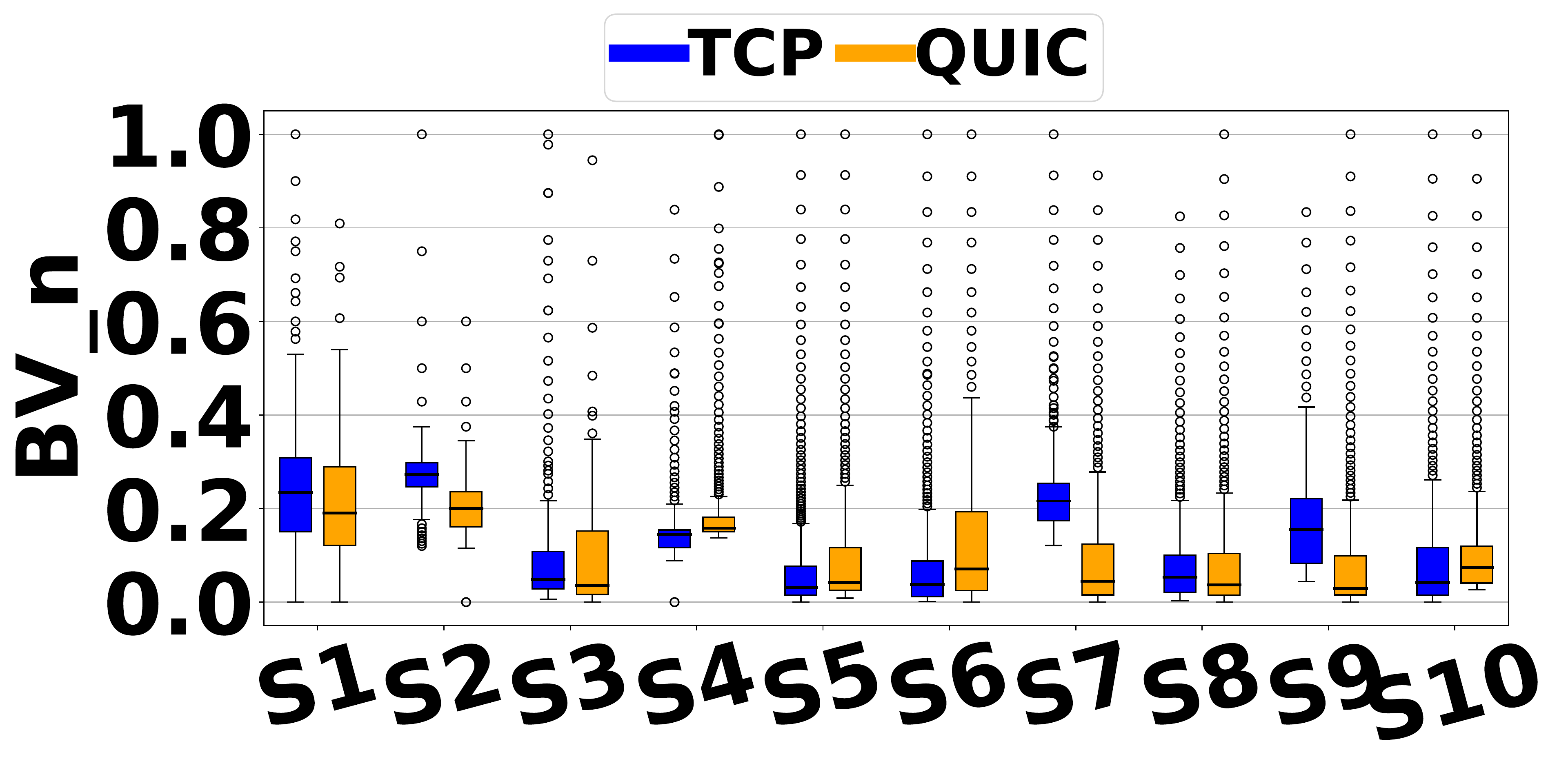}}
\subfloat[]{\includegraphics[width = 0.25\textwidth]{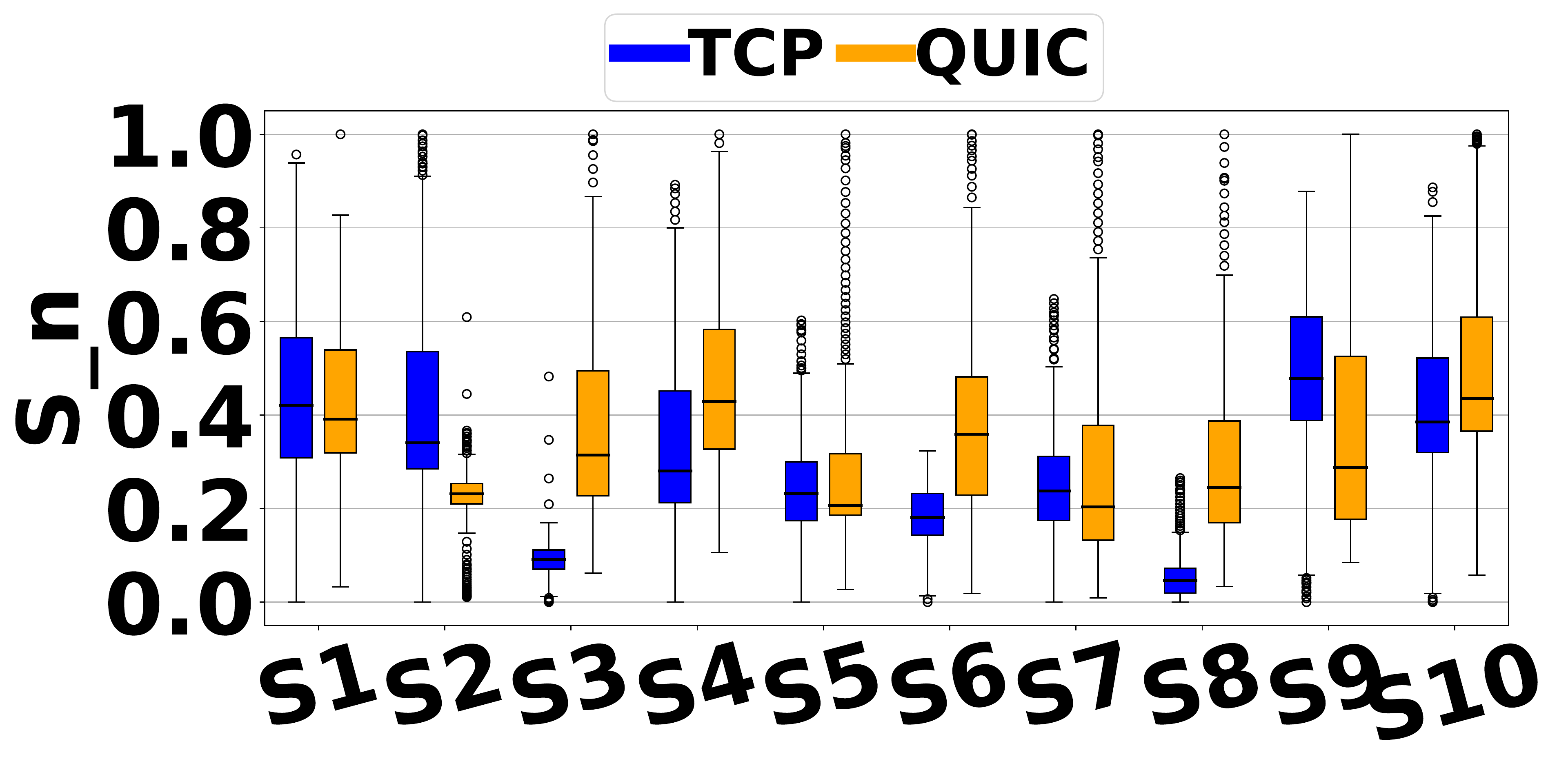}} 
\caption{Normalized QoE ($Q\_n$), average bitrate ($B\_n$), average bitrate variation ($BV\_n$) and average stall ($S\_n$) distribution of $10$ sample streaming pairs.}
\label{individual-para}
\end{figure}

Fig.~\ref{individual-para}(b), (c), and (d) show the distribution of average bitrate, average bitrate variation, and average stall for $10$ sample streaming session pairs\footnote{A session pair consists of two streaming sessions -- one over QUIC-enabled browser and the other over QUIC-disabled browser (pure TCP streams) while keeping all other parameters, including traffic shaping over the time, unchanged.}. Out of these $10$ session pairs, we considered $4$ from the DL category (check Section~\ref{sec:expt-setup}), namely S1 to S4, and the remaining $6$ from the DVL category, namely S5 to S10. Fig.~\ref{individual-para}(a) shows that in terms of the overall normalized QoE, TCP streams work better than QUIC-enabled streams for session pairs S3, S5, S7, S8, and S9 (all with $p < 0.05$). However, if we look into individual QoE parameters, in all of the above five session pairs, TCP streams provide a better average bitrate compared to QUIC-enabled streams (with $p < 0.05$). On the contrary, we observe that S3, S4, S6, S8, and S10 yields less average stall over TCP streams than QUIC-enabled streams (with $p < 0.05$). In particular, the cases for S4 and S10 are interesting when TCP-streams result in less stall than QUIC-enabled streams, but at the same time play the video over less average bitrate, thus resulting in lower average QoE compared to TCP streams. On the other hand, in S5 and S7, a video playback over TCP streams results in more stall than QUIC-enabled streams and utilizes a higher average bitrate (statistically significant with $p < 0.05$), thus ensuing better QoE overall. In a nutshell, we observe that individual video performance in terms of different QoE parameters differs significantly over different video sessions, and a QUIC-enabled stream does not always ensure a better performance. These observations also follow our observations on the hypothesis testing over individual QoE parameters for all the session pairs, as shown earlier in Fig.~\ref{hypothesis-result}(b).

\subsection{Does Bandwidth Pattern Impact QUIC-enabled Streams?}
\begin{figure}[t]
\subfloat[]{\includegraphics[width = 0.25\textwidth]{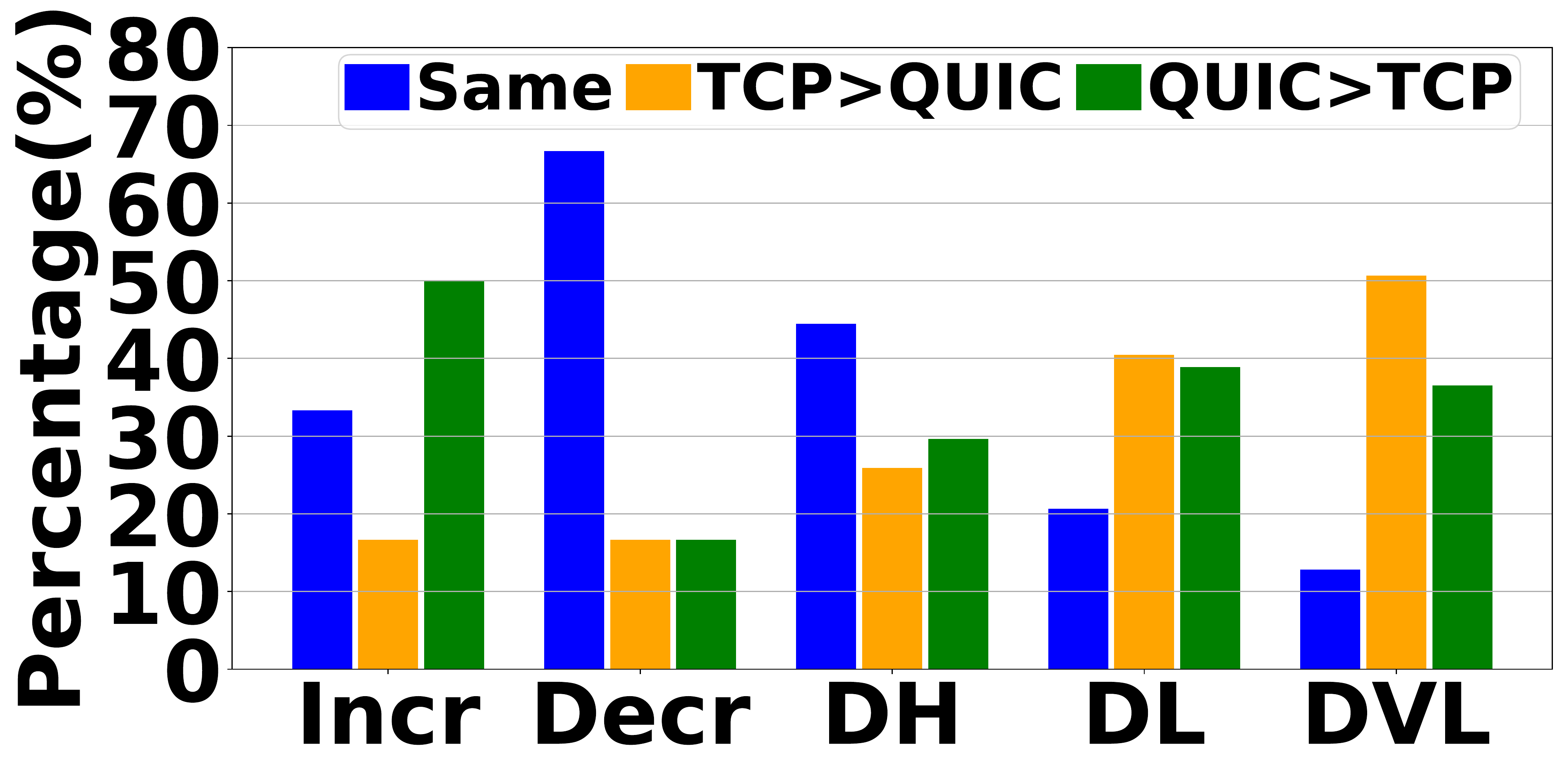}}
\subfloat[]{\includegraphics[width = 0.25\textwidth]{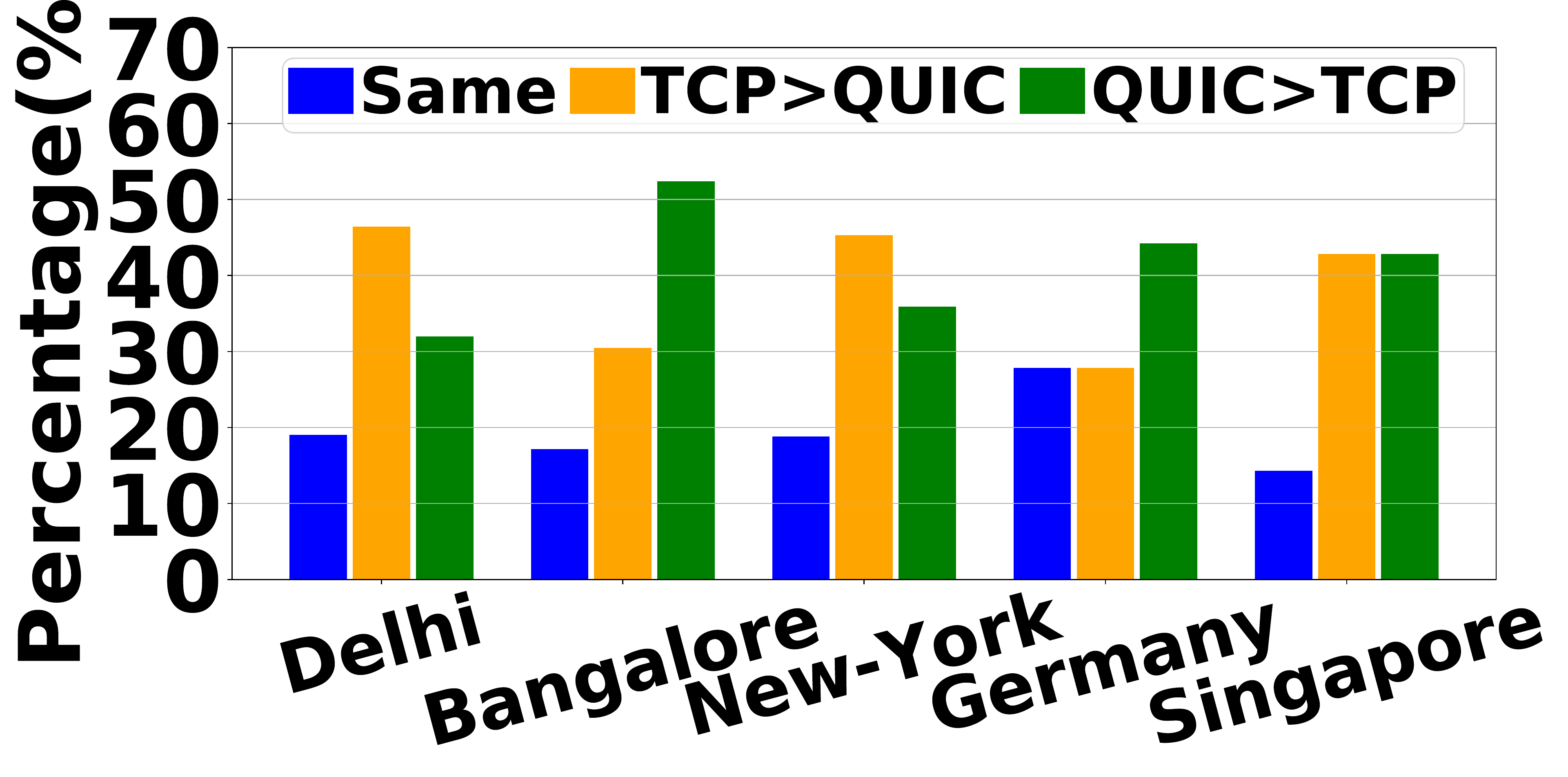}}
\subfloat[]{\includegraphics[width = 0.25\textwidth]{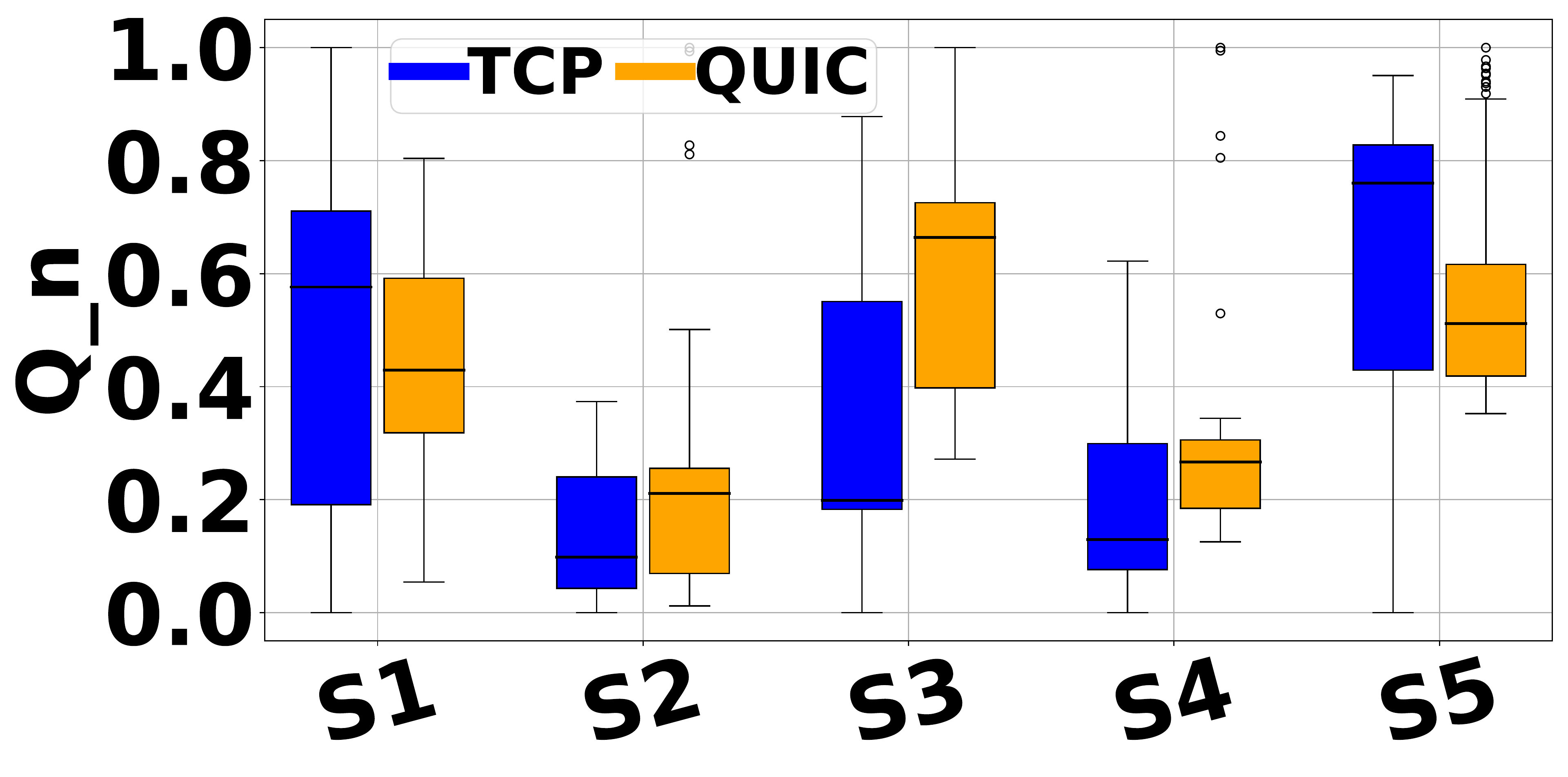}}
\subfloat[]{\includegraphics[width = 0.25\textwidth]{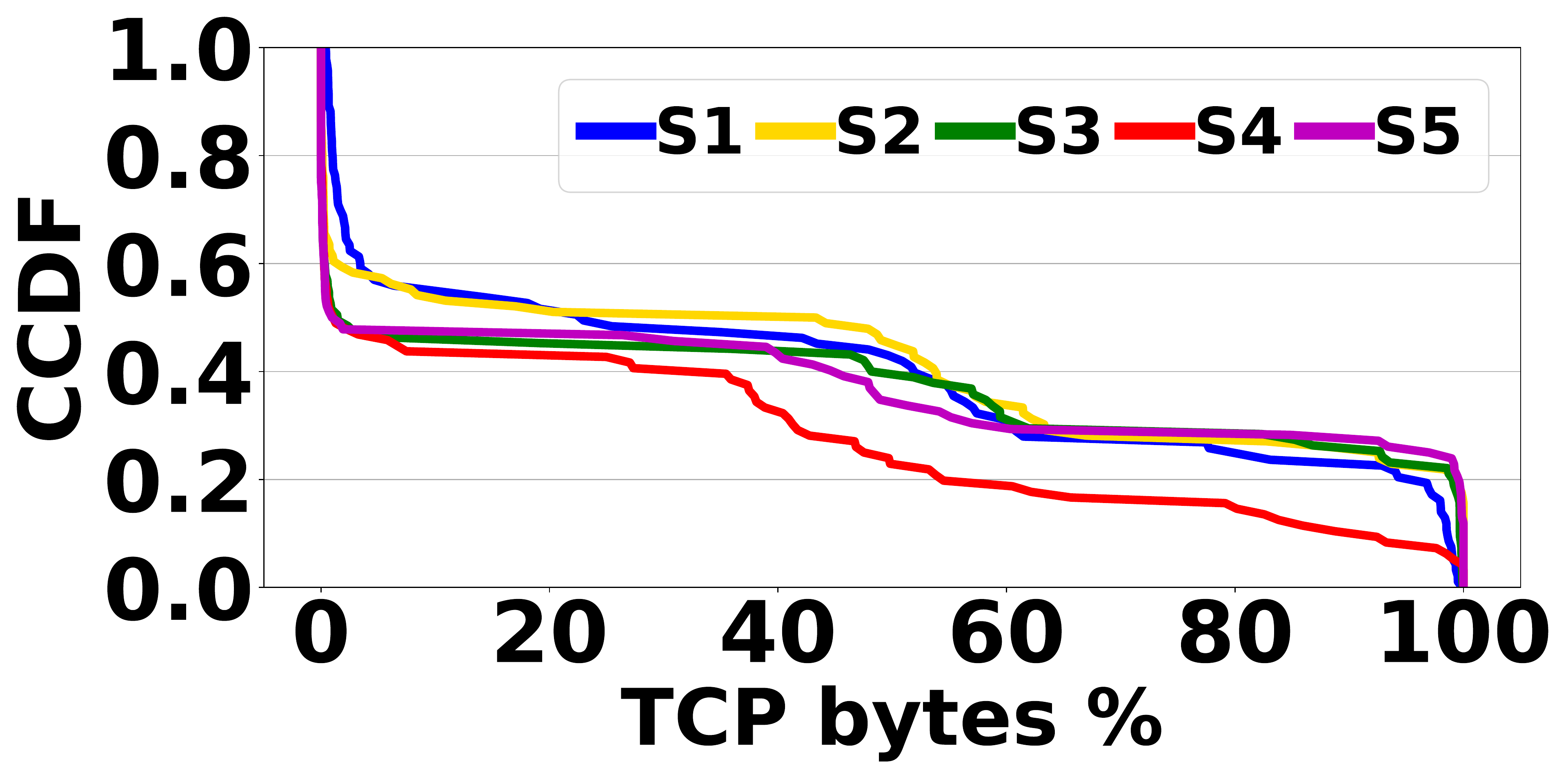}}
\caption{(a) Hypothesis testing results across various bandwidth patterns and (b) Hypothesis testing results across various geographical locations (c) Normalized QoE ($Q\_n$) distribution for same video ran multiple times and (d) TCP bytes $\%$ graph for same video ran multiple times}
\label{hypothesis-bandwidth-location}
\end{figure}

We collected data for different bandwidth patterns through the traffic shaper (Section~\ref{sec:expt-setup}); here, we performed the hypothesis testing and two-tail test to determine what percentage of cases TCP streams perform better than QUIC-enabled streams and vice-versa under different traffic shaping scenarios. From Fig.~\ref{hypothesis-bandwidth-location}(a), we observe that for the scenarios under DL and DVL, TCP streams perform better than QUIC-enabled streams for more than $40\%$ of the video playback scenario pairs. Additionally, we observe that in DVL, most of the videos were played at the minimum quality, i.e., at $144$p. In this case, the instances of quality drops are significantly less; instead, we observe more rebuffering events. For DVL, as there is a scope of a quality drop to counter poor bandwidth, there are more quality drop events along with rebuffering events. Correlating these observations with the observation that we had earlier in Fig.~\ref{fallback-percent}(b), we see that percentage of TCP traffic is more over the QUIC-enabled streams for DL and DVL. This indicates that instances of fallbacks are more in case of such low bandwidth scenarios, thus raising the question -- \textit{does fallback causes poor performance in QUIC-enabled streams for such scenarios?} In other words, \textit{should we disable QUIC over a browser when the device is connected with a low-bandwidth network?}

\subsection{Are Our Observations Consistent Across Geographical Locations?} 
To answer this question, we ran the videos on different geographical locations namely Delhi, Bangalore, New York, Germany and Singapore using digital ocean machines (details in Section~\ref{sec:expt-setup}). Fig~\ref{hypothesis-bandwidth-location}(b) shows the result of hypothesis testing and two-tail test across individual streaming session pairs collected over these five geographical regions. The figure indicates that QUIC-enabled streams performed better than TCP streams for more number of cases in Bangalore and Germany. However, across all the five geographical regions, there are more than $30\%$ of the scenarios when TCP streams yielded better application QoE compared to QUIC-enabled streams. Again, correlating this figure with Fig.~\ref{fallback-percent}(c), we can see that Delhi and New York have more instances of TCP traffic within the QUIC-enabled streams, compared to other locations; Fig~\ref{hypothesis-bandwidth-location}(b) shows that TCP streams are good for these two cities (Delhi and New York) compared to QUIC-enabled streams. Therefore, we again observe an indication that fallbacks in QUIC-enabled streams may hinder the application QoE rather than helping!

\subsection{Are the Above Observations Specific to a Particular Video Genre?}
One might think that QUIC-enabled streams perform poorly for some particular video genre, like for action or entertainment videos that might have frequent changes in scenes and hence need frequent changing in bitrates for variable bitrate encoding. To answer this research question, we streamed an entertainment video $32$ times over YouTube at DVL, however on different days or at different times of the day. Fig.~\ref{hypothesis-bandwidth-location}(c) shows the QoE distribution of the sample video for five randomly selected sample iterations. We observe that for S1, TCP stream  yields a better performance ($p<0.05$) compared to QUIC-enabled stream. However, for S3, the QUIC-enabled stream resulted in a better performance ($p<0.05$) compared to TCP stream. Fig.~\ref{hypothesis-bandwidth-location}(d) shows the CCDF of TCP bytes (percentage) transferred. For $50$\% times, S1 transfers about $10$\% or more bytes over TCP and S3 transfers about $1$\% bytes over TCP. This also indicates that fallback makes a QUIC-enabled stream to suffer more compared to its pure TCP counterpart. 
\subsection{Are the Above Observations Consistent Across Time?}
\begin{wrapfigure}{r}{0.3\textwidth}
\includegraphics[width = 0.3\textwidth]{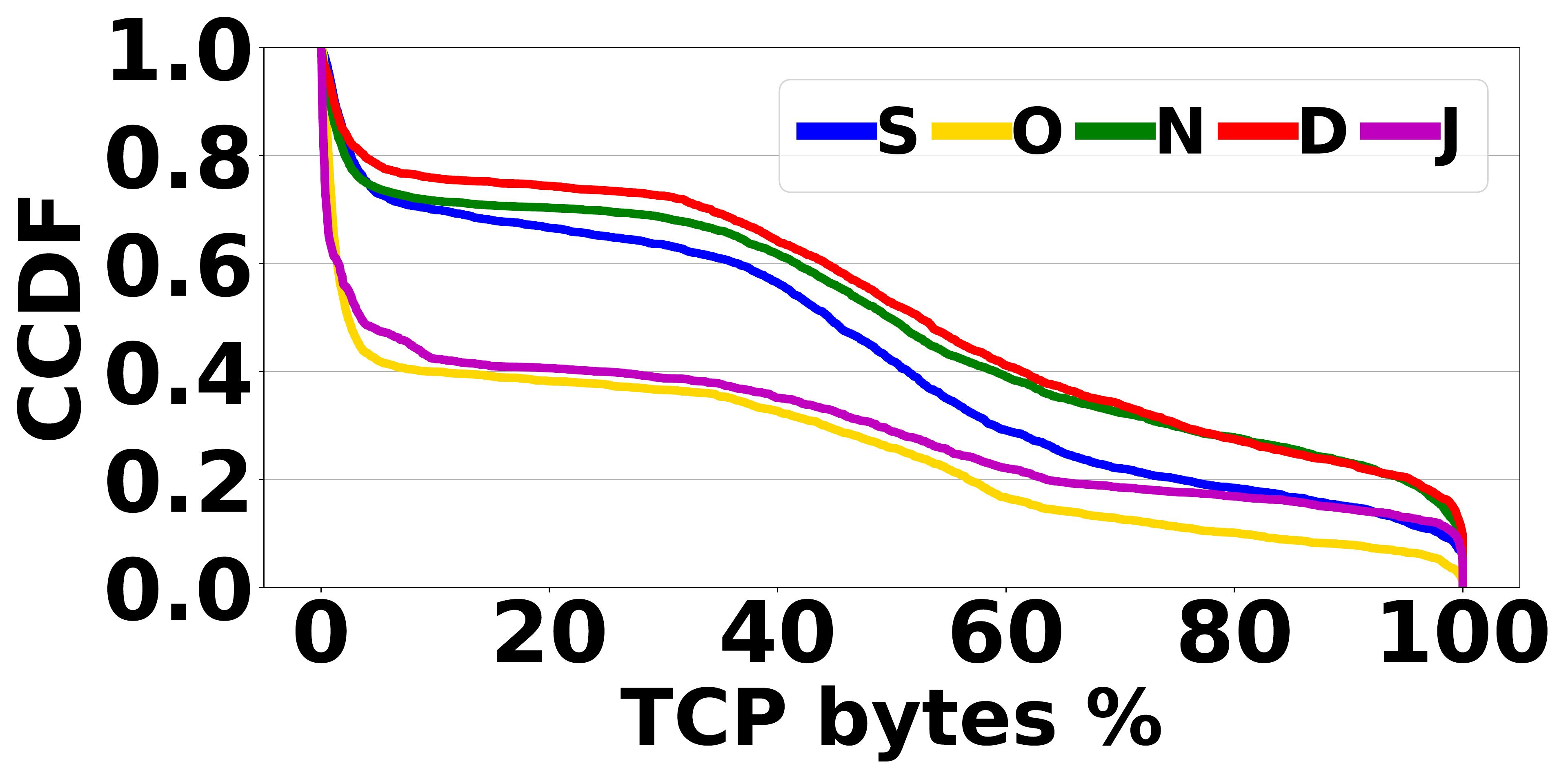}
\caption{Percentage of fallback, i.e., percentage of bytes transferred over TCP in a QUIC-enabled stream across months (S=September, O=October, N=November, D-December, J=January)}
\label{fallback-month}
\end{wrapfigure}
We wondered if these fallbacks are due to some temporary issues. Hence, we look at the instances of fallbacks over months.  Fig.~\ref{fallback-month} shows the complementary CDF (CCDF) of the percentage of TCP bytes transferred over $5$ months, namely September, October, November, December, and January. We noticed significant fallback across all months. Interestingly, we observe that $40$\% times, more than $9$\% and $24$\%  are transferred over TCP for October and January respectively. Whereas, for September, November, and December $40$\%  times, more than $51$\% , $59$\%, and $61$\% are transferred over TCP respectively. 
Hence, we conclude that our observations are consistent across time. 

\subsection{Summary Of Observations} 
The current and the previous sections analyzed and compared the performance of TCP streams and QUIC-enabled streams in terms of YouTube video streaming QoE under different traffic shaping and over various geographical regions. We have a few consistent observations across all these analyses. \textbf{First}, QUIC-enabled streams may yield similar or better performance compared to TCP streams in the long run (when we collectively analyze over large samples). However, in the short run, i.e., for individual streaming sessions, a video may suffer if QUIC is enabled over the browser. \textbf{Second}, a QUIC-enabled YouTube video stream not only carries QUIC traffic but falls back to TCP whenever the 1-RTT/0-RTT connection establishment fails. Instead, Google's QUIC initiates a streaming session with TCP and migrates to QUIC if a parallel QUIC connection is successful. \textbf{Third}, under different bandwidth patterns locations, we observe a significant number of instances where pure TCP streams outperform QUIC-enabled streams. Most interestingly, in many of those instances, the QUIC-enabled streams carried more TCP traffic than QUIC traffic. These observations make us think whether such fallbacks are one of the primary root causes behind the lagging performance of Google's QUIC. To answer this question, we performed a thorough statistical analysis, particularly correlation and causality analysis, over all the samples that we have collected.   

\section{Correlation and Causality Analysis}
\label{sec:corr}
We next perform statistical correlation and causality analysis on the QoE parameters and the captured fallback events from the QUIC-enabled streams. From the hypothesis testing, as reported in Section~\ref{sec:analysis1} (Fig.~\ref{hypothesis-result}), we consider the cases when a TCP stream works better than the QUIC-enabled stream in video playback session and analyze whether fallback was one of the primary causes behind the poor performance of QUIC-enabled streams. Modeling correlation and causality is particularly challenging in this case because we have to work with three different time axes -- the \textit{real-time} (the time of the global clock when a video frame is rendered), the \textit{playback time} (the relative time of a particular video frame with respect to the video start time) and the \textit{download time} (the time when a video frame has been downloaded). 

\subsection{System Model and Problem Statement} 
Let $t$ be the global clock time. Let $f_t$ be the video frame that has been rendered over the YouTube client at time $t$. We assume that $\tau(f_t)$ be the playback time for the frame $f_t$ and $\sigma(f_t)$ be the download time for the frame $f_t$. It can be noted that $\tau(f_t) \leq t$ as the frame $f_t$ can be rendered either at its original playback time $\tau(f_t)$ (if there is no rebuffering or video stall) or after that (if there is a stall before rendering the frame). Also, $\sigma(f_t) < t$, as $f_t$ needs to be downloaded before it is rendered. Further, there would be a rebuffering if $\sigma(f_t) > \tau(f_t)$, i.e. the video frame $f_t$ is downloaded after its original playback time. 

Let $\mathcal{B}_v(t)$ and $\mathcal{S}_v(t)$ denote the perceived bitrate and video rebuffering at time $t$ during a video playback session $v$. Here $t$ is the real time (in msec) of the system, and not the playback time. Let $e(q)$ be the encoded quality level at which the video is being played; for a video quality level $q$, we encode it using the mapping $q \rightarrow e(q)$ as follows: 144p $\rightarrow$ 1, 240p $\rightarrow$ 2, 360p $\rightarrow$ 3, 480p $\rightarrow$ 4, and 720p $\rightarrow$ 5. Let $q_t$ be the quality level at which the video frame $f_t$ is being played, $q_t^{prev}$ be the quality level at which the video was being played just before it switched to the quality level $q_t$ (so, there is a quality switch $q_t^{prev} \rightarrow q_t$), and $t_{dur}(q_t)$ be the duration of playback after it switched to $q_t$ from $q_t^{prev}$. Then, we model $\mathcal{B}_v(t)$ as the time series, $\mathcal{B}_v(t) = (e(q_t^{prev}) - e(q_t)) \times t_{dur}(q_t)$. Similarly, we model $\mathcal{S}_v(t) = t - \tau(f_t)$, following Fig.~\ref{stall-comp} (in Sec.~\ref{sec:expt-setup}).   

Finally, let us assume that $\mathcal{F}_v(t)$ denotes the fallback at time instance $t$ for a QUIC-enabled stream corresponding to the video playback session $v$. Let $\mathbb{T}(t)$ and $\mathbb{Q}(t)$ denote the amount of TCP and QUIC traffics, respectively, in the QUIC-enabled stream up to time $t$. Then we model $\mathcal{F}_v(t) = \frac{\mathbb{T}(t)}{(\mathbb{T}(t) + \mathbb{Q}(t))}$.\\

\noindent\textbf{Problem Definition:} The objective of this section is to statistically analyze whether $\mathcal{B}_v(t)$ and $\mathcal{S}_v(t)$ correlate with $\mathcal{F}_v(t)$, and if they correlate, then whether $\mathcal{F}_v(t)$ causes either $\mathcal{B}_v(t)$ or $\mathcal{S}_v(t)$ or both.\\ 

\noindent\textbf{Challenges:} Let $f_t$ be the video frame being rendered at time $t$; so $\mathcal{B}_v(t)$ and $\mathcal{S}_v(t)$ capture whether $f_t$ experiences any quality drop or rebuffering. However, as we mentioned earlier, $\sigma(f_t) < t$; therefore $\mathcal{F}_v(t)$ does not necessarily capture what happened to the transport when $f_t$ is being downloaded. Rather, the impact of fallback might get visible on the application several times after the fallback has actually happened. Therefore, there is likely to be a \textit{lag} between the two time series $\mathcal{F}_v(t)$ and $\mathcal{B}_v(t)$, as well as between $\mathcal{F}_v(t)$ and $\mathcal{S}_v(t)$. We need to consider the effect of this lag during the correlation and the causality analysis.

\subsection{Correlation Analysis}
\label{cc-about}
\begin{figure}[!t]
\subfloat[]{\includegraphics[width = 0.25\textwidth]{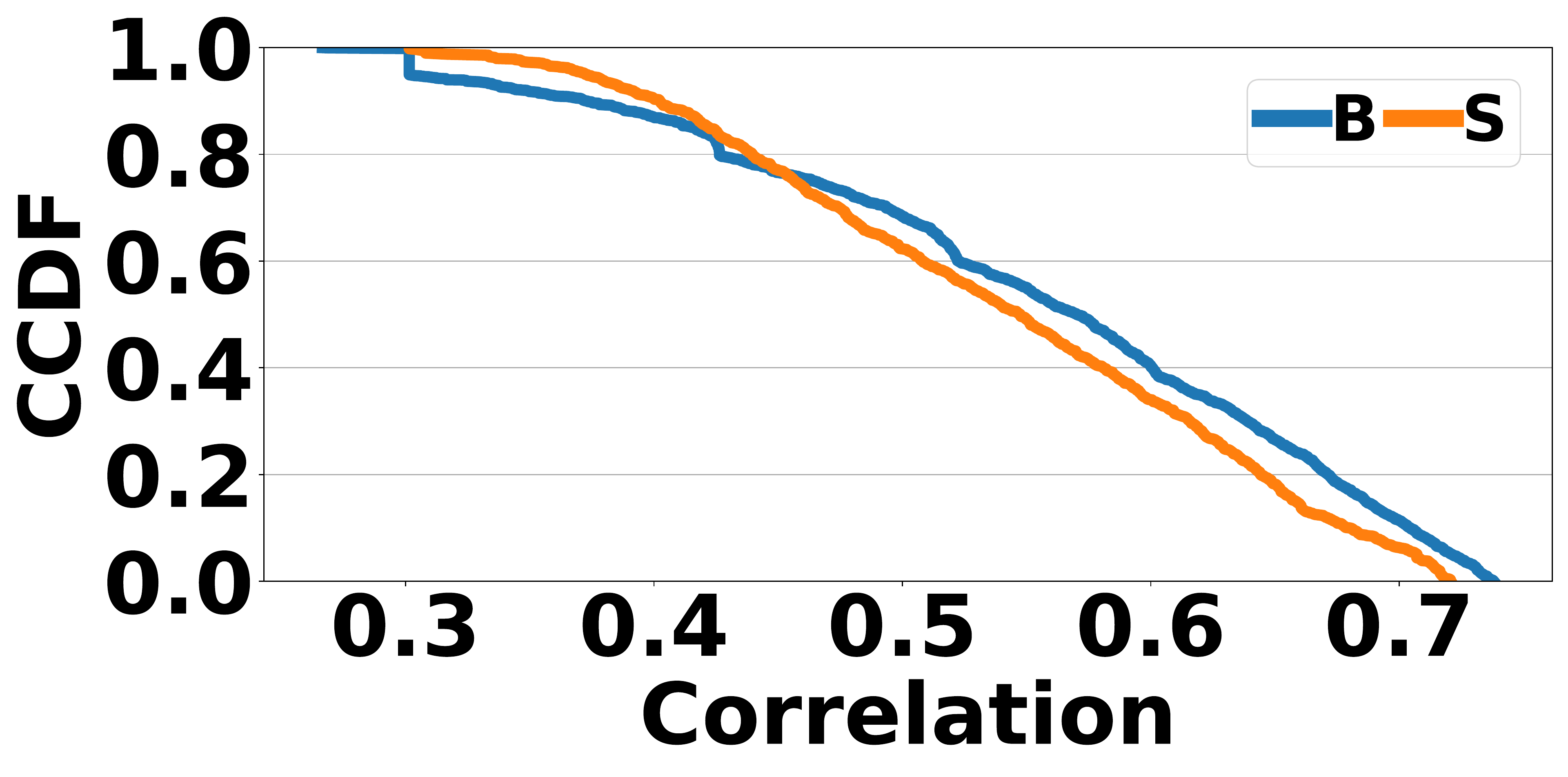}} 
\subfloat[]{\includegraphics[width = 0.25\textwidth]{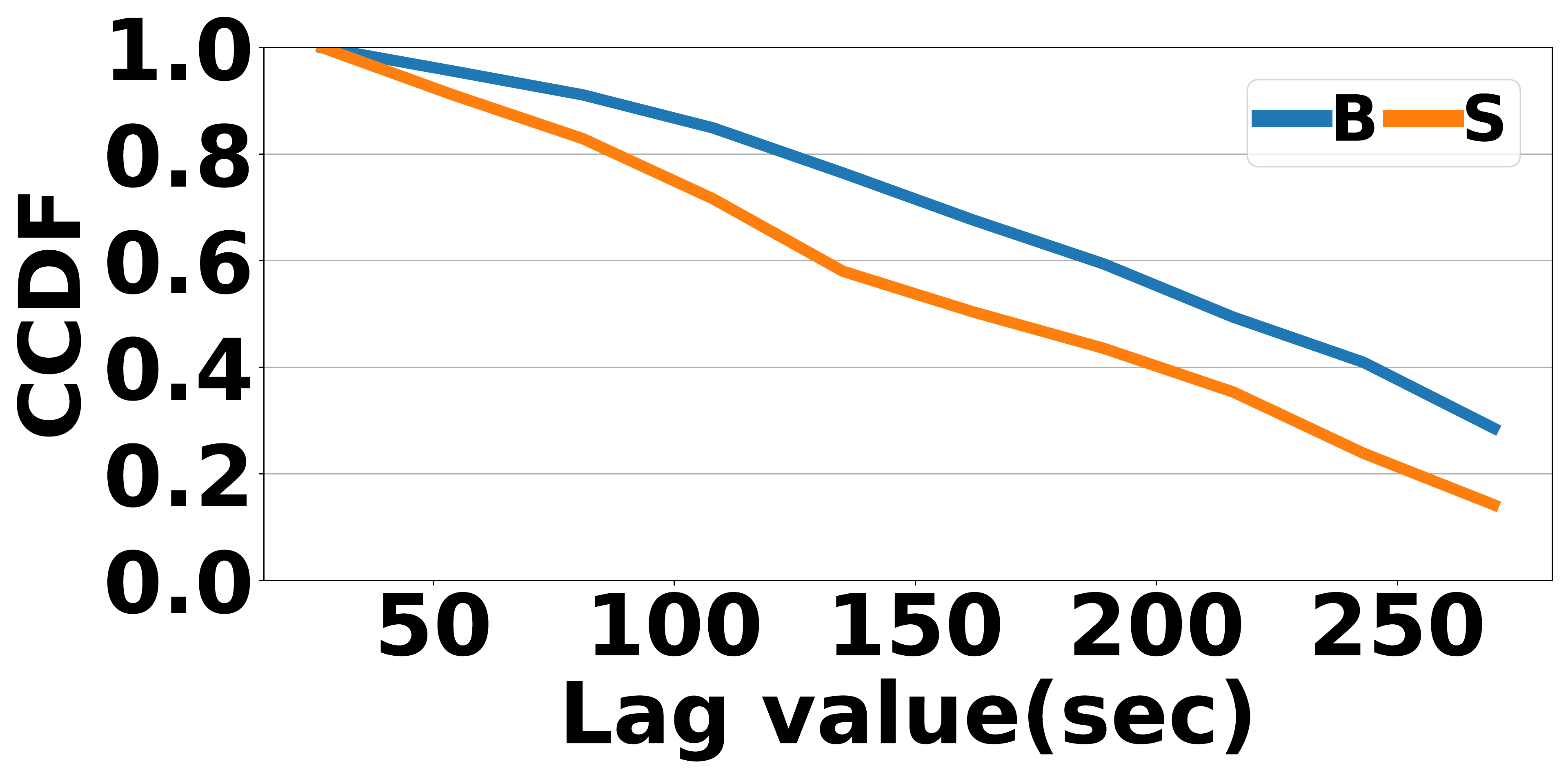}}
\subfloat[]{\includegraphics[width = 0.25\textwidth]{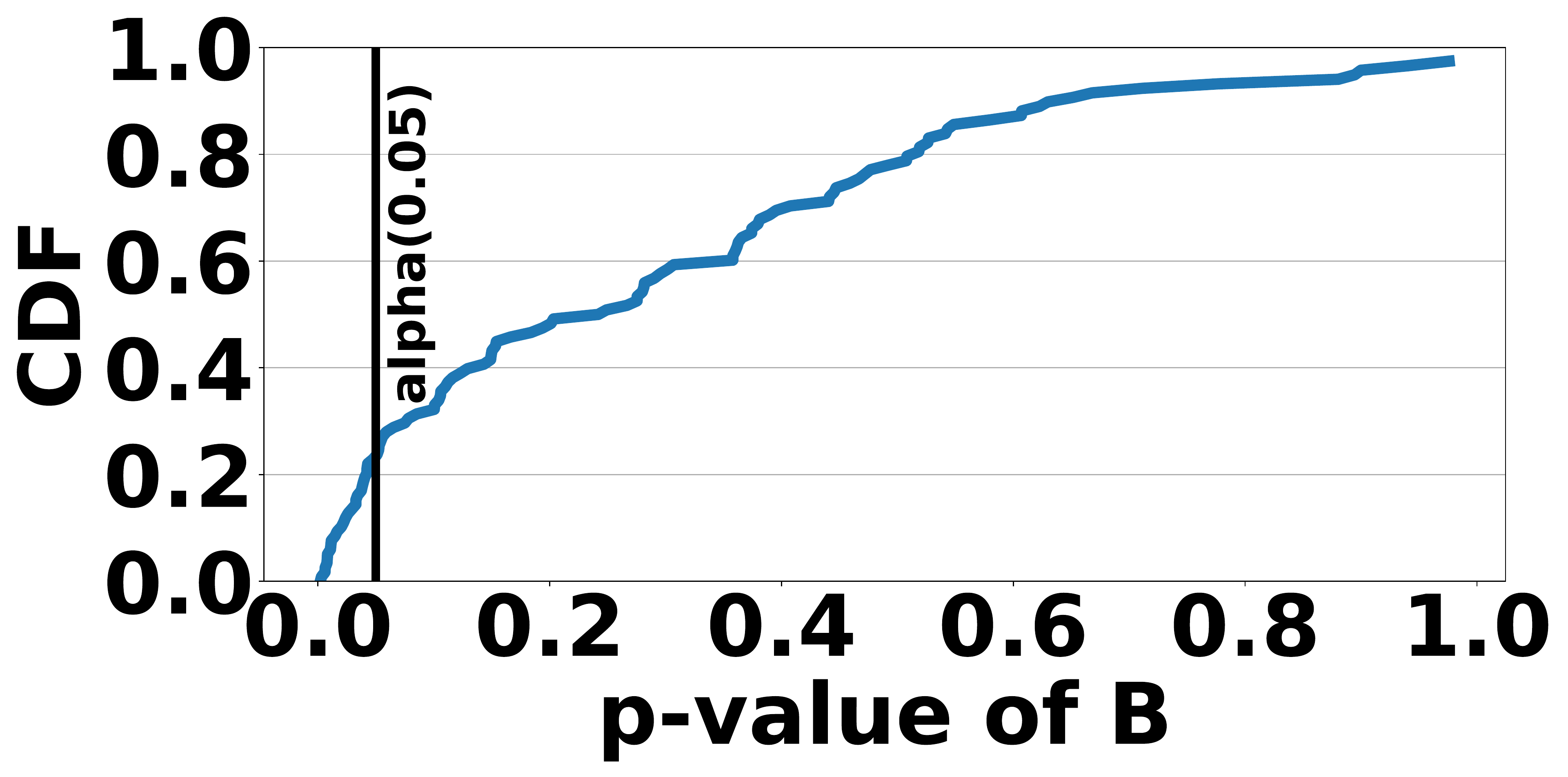}} 
\subfloat[]{\includegraphics[width = 0.25\textwidth]{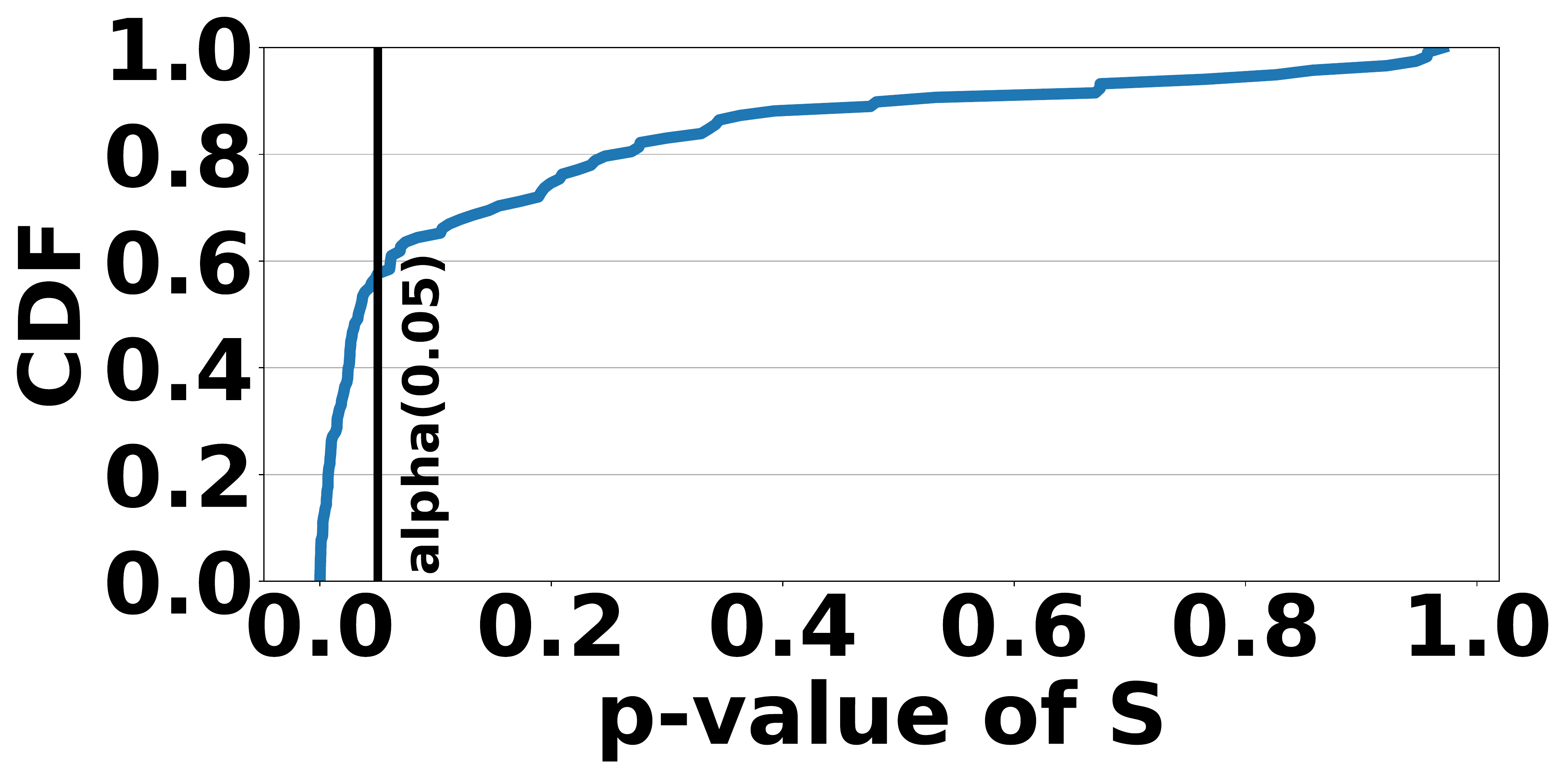}}
\caption{(a) Cross correlation between fallback $\mathcal{F}$ and quality drop $\mathcal{B}$, (b) Cross correlation between fallback $\mathcal{F}$ and stalling $\mathcal{S}$, (c) Granger Causal Analysis between fallback $\mathcal{F}$ and quality drop $\mathcal{B}$, (d) Granger Causal Analysis between fallback $\mathcal{F}$ and stalling $\mathcal{S}$}
\label{cross-corr}
\end{figure}

We compute the cross-correlation between the two time-series data $\mathcal{F}_v(t)$ and $\mathcal{B}_v(t)$ as follows. 
\begin{equation}
    (\mathcal{F}_v*\mathcal{B}_v)(\mu_b) = \int\limits_{-\infty}^{\infty}\overline{\mathcal{F}_v(t)}\mathcal{B}_v(t+\mu_b)dt
\end{equation}
where $\overline{\mathcal{F}_v(t)}$ is the complex conjugate of $\mathcal{F}_v(t)$ and $\mu_b$ is the corresponding lag. We compute the $\mu_b$ for which we obtain the maximum cross-correlation ($\mu_b^{max}$). Similarly, we compute the cross-correlation between $\mathcal{F}_v(t)$ and $\mathcal{S}_v(t)$ as follows. \begin{equation}
    (\mathcal{F}_v*\mathcal{S}_v)(\mu_s) = \int\limits_{-\infty}^{\infty}\overline{\mathcal{F}_v(t)}\mathcal{S}_v(t+\mu_s)dt
\end{equation}
where $\mu_s$ is the corresponding lag. Similar to the above, we compute the $\mu_s$ for which we obtain the maximum cross-correlation ($\mu_s^{max}$). We next plot the CCDF distributions of correlation and the lag values in Fig.~\ref{cross-corr}. 

Fig.~\ref{cross-corr}(a) shows the correlation between fallback ($\mathcal{F}_v(t)$) and quality drop ($\mathcal{F}_v(t)$) as well as stalling ($\mathcal{S}_v(t)$). Similarly, Fig.~\ref{cross-corr}(b) shows the lag values $\mu_b^{max}$ and $\mu_s^{max}$. We observe that about $50\%$ of videos positively correlate with a correlation coefficient of $0.55$ or higher for both cases. Therefore, we get an indication that for several video streaming sessions over QUIC-enabled streams, fallback incidents show a good correlation with the perceived QoE. However, correlation does not apply causality; therefore, we proceed with modeling causality between these time series distributions. While doing so, we observe from Fig.~\ref{cross-corr}(b) that for both $\mathcal{B}_v(t)$ and $\mathcal{S}_v(t)$, the lag values across different streaming sessions are distributed within $250$sec. We use this information in the causality analysis, as discussed next.

\subsection{Causality Analysis}
We apply linear Granger causality~\cite{Gc} on the collected video streaming sessions, which is a statistical hypothesis test that determines whether one time series can be used to forecast another. We use a linear model to test Granger causality in the mean. More specifically, for a video session $v$, we consider two test cases -- Test $B_v$ and Test $S_v$. We consider the null hypothesises to be ``\textit{$\mathcal{F}_v(t)$ does not Granger causes $\mathcal{B}_v(t)$}'' and ``\textit{$\mathcal{F}_v(t)$ does not Granger causes $\mathcal{S}_v(t)$}'', respectively for the two tests $B_v$ and $S_v$. Granger causality assumes the time series to be stationary; if the time series is not stationary, the causality test can be performed using a transformed series of first-order or second-order differentiation. 

Mathematically, Granger causality tests the null hypothesis as follows. Considering the test $B_v$ with an assumption that both $\mathcal{F}_v(t)$ and $\mathcal{B}_v(t)$ are stationary time series, $\mathcal{B}_v(t)$ is first predicted using univariate auto-regression over the lagged values of $\mathcal{B}_v(t)$;
\begin{equation}
    \mathcal{B}_v(t) = a_0 + a_1\mathcal{B}_v(t-1) + \cdots + a_m\mathcal{B}_v(t-m) + E_i
\end{equation}
where $m$ is the maximum lag value and $E_i$ is the prediction error. Then, this auto-regression is augmented using the lagged values of $\mathcal{F}_v(t)$ as follows.
\begin{equation}
    \mathcal{B}_v(t) = a_0 + a_1\mathcal{B}_v(t-1) + \cdots + a_m\mathcal{B}_v(t-m) + b_x\mathcal{F}_v(t-x) + \cdots + b_q\mathcal{F}_v(t-q) E_i
\end{equation}
where $x$ and $y$ are the minimum and the maximum lag length for which the lag values of $\mathcal{F}_v$ is individually significant in predicting $\mathcal{B}_v$ according to their t-statistics and they collectively add explanatory power to the regression according to an F-test. The null hypothesis is accepted if no lag values of $\mathcal{F}_v$ is retained in the above regression. The test $S_v$ can also be performed in a similar way. 

We perform the above analysis over our dataset as follows. For a video session $v$ over the QUIC-enabled stream, we first use Kwiatkowski–Phillips–Schmidt–Shin (KPSS) test~\cite{Kpss} to check whether the time series $\mathcal{F}_v$, $\mathcal{B}_v$ and $\mathcal{S}_v$ is stationary. If at least one of them is non-stationary, we transform all three time series using first-order differentiation and apply the KPSS test again. Even if one of them is found to be non-stationary, we again perform second-order differentiation on all three time series and perform the KPSS test. A scenario is marked as unsuitable for the Granger causality test if at least one of them is non-stationary even after second-order differentiation. We found that around $2\%$ of all the scenarios are not suitable for the Granger causality test. For the remaining scenarios, we perform the Granger causality test either on the time series or their transformation and collect the individual $p$-values for the tests for both test $B_v$ and test $S_v$. 

In Fig.~\ref{cross-corr}(c) and (d), we plot the CDF of $p$-values obtained over the above two tests. We reject the null hypothesis ``\textit{$\mathcal{F}_v(t)$ does not Granger cause $\mathcal{B}_v(t)$}'' for  $20$\% cases (obtaining a $p$-values $<0.05$). Similarly, we reject null hypothesis ``\textit{$\mathcal{F}_v(t)$ does not Granger cause $\mathcal{S}_v(t)$}'' for $60$\% cases. We computed union of these two series and that turns out to $70\%$. Hence, we can conclude $\mathcal{F}_v(t)$ Granger causes either $\mathcal{B}_v(t)$ or $\mathcal{S}_v(t)$ for $70$\% of the scenarios where we observed that a TCP stream works better than QUIC-enabled streams. Interestingly, as we discussed earlier, video rebuffering or stalls are more prevalent under DL and DVL; further, much of the fallbacks happen when the bandwidth is low. Therefore, a QUIC-enabled browser may affect the streaming performance more by causing additional stalls than reducing the connection setup latency using 1-RTT or 0-RTT connection establishment process.

\section{Discussion}
Our paper suggests that it is better to turn off QUIC when there is a possibility that a QUIC connection establishment might not succeed. During a QUIC-enabled session, migrating to QUIC from TCP and then again falling back to TCP due to a broken QUIC connection indeed hurts the application performance. We observed that this ping pong among TCP and QUIC is particularly alarming during a streaming session over a low-bandwidth network. However, the sheer variance in the real world makes it impossible to measure precisely. Though we have tried to emulate the real world to understand what happens underneath when an application like YouTube streaming suffers over a QUIC-enabled session, our observations are also limited to a few use cases.\\

\noindent\textbf{(1) Observations are valid for Google's QUIC only.} The analysis presented in this paper is specific to Google's QUIC for YouTube streaming. QUIC has several implementations, and different implementations of QUIC use different configurations, as mentioned in~\cite{dissecting}. Indeed, our findings may be specific to a set of configurations that Google uses internally. It can be noted that we have not made any attempt to change the QUIC implementation within the Chrome browser. Accordingly, Facebook's QUIC or Cloudflare's QUIC might inhibit a different observation. Nevertheless, given that Google was the original developer of the QUIC protocol, it is worthy of understanding the behavior of Google's QUIC from the perspective of a well-established commercial end-point like YouTube streaming.\\

\noindent\textbf{(3) There are other notions of causality.} We have used the Granger causal model in our analysis, which utilizes a prediction pipeline based on linear regression to check whether one time series can be used to predict another. However, there are other notions of causality, such as based on \textit{Average Treatment Effect}~\cite{kovandzic2013estimating}, \textit{Twin Study}~\cite{martin1978power}, \textit{Propensity Score Matching}~\cite{caliendo2008some}, which might provide a different inference from what we have observed in this study. However, the important notion of our study is that we modeled fallback using the volume of TCP packets transmitted over a QUIC-enabled stream, which also indirectly indicates the network condition. Typical ABR algorithms use the measure of network condition as one of the prime factors to decide the bitrate of a segment. Therefore, this notion of fallback can be an ideal measure to predict the QoE drop over a streaming application, which we have utilized through the Granger causality model. \\ 

\noindent\textbf{(3) Linear causality might not justify everything.} We have used linear Granger's causality; although it shows that fallbacks cause QoE drops for around $70\%$ of the cases, there might be instances where nonlinear causality might justify the impact. However, nonlinear Granger's causality needs complex estimation based on deep learning. Independent video sessions have fixed duration, and our analysis is restricted in comparing individual video sessions over TCP and QUIC-enabled streams. It isn't easy to train a deep model using the data from a single streaming session only. Therefore, we have been restricted to linear analysis only. Interestingly, with stationary transformation, linear Granger causality could justify around $70\%$ of the observations, which is a considerable proportion of all our observations.\\

\noindent\textbf{(4) Fallback might not be the only cause.} Our analysis indicates that fallback is a prime reason behind the poor performance of a YouTube streaming over a QUIC-enabled session. However, there are instances where QUIC-enabled streams perform poorly, even in a high bandwidth scenario when there is no fallback. Therefore, further research is required to find out what can be the other reasons that affect the performance of a QUIC stream over in-the-wild Internet.\\ 
 
Finally, we believe that an analysis over more than $2600$ streaming hours of video data can provide a reliable indication about the root cause of QUIC's poor performance, as reported in much of the previous literature. An interesting observation is that QUIC does not differentiate a failure caused by a middlebox and the failure due to network congestion, which needs a thorough investigation to support a stable performance of QUIC over the Internet.  
 

\section{Conclusion}
In this paper, we analyze the performance of a QUIC-enabled stream for YouTube video streaming over varying bandwidth patterns and across different geographical locations. We collected the dataset by streaming $2046$ streaming sessions, covering more than $2600$ streaming hours. We conducted statistical testing and found that a QUIC-enabled stream is not a winner many of the times. This is especially true for low bandwidth. Further, our observation is consistent across other geographical locations and videos. Interestingly, we observe that many of the QUIC connections fall back to TCP in poor bandwidth. While this fallback option provides sustainability across middleboxes, but QUIC protocol does not distinguish between failures caused by middleboxes vs. those caused due to network congestion. We finally prove statistically that QUIC's fallback option to sustain across middleboxes is one of the root causes for such poor performance of a QUIC-enabled stream.

Given that Google was the original developer of the QUIC protocol, this study suggests a revisit of QUIC's implementation for its sustainability over the Internet. Our study does not contradict QUIC's superiority over TCP; QUIC certainly has features like stream multiplexing and 0-RTT connection establishment that can help an application boost its performance. However, the protocol has to sustain over today's network that is not friendly for many network protocols, particularly those like UDP that can open a potential threat window for an organization. The 0-RTT connection establishment cannot help if the connection establishment packets need to be retransmitted multiple times. Accordingly, it is essential to design policies to support QUIC-enabled streams through middleboxes while ensuring QUIC's performance benefits. 



\bibliographystyle{ACM-Reference-Format}
\bibliography{bibliography}
\newpage
\section*{Appendix}
\label{sec:appendix}
\begin{figure}[H]
\centering
\includegraphics[width =0.95\textwidth]{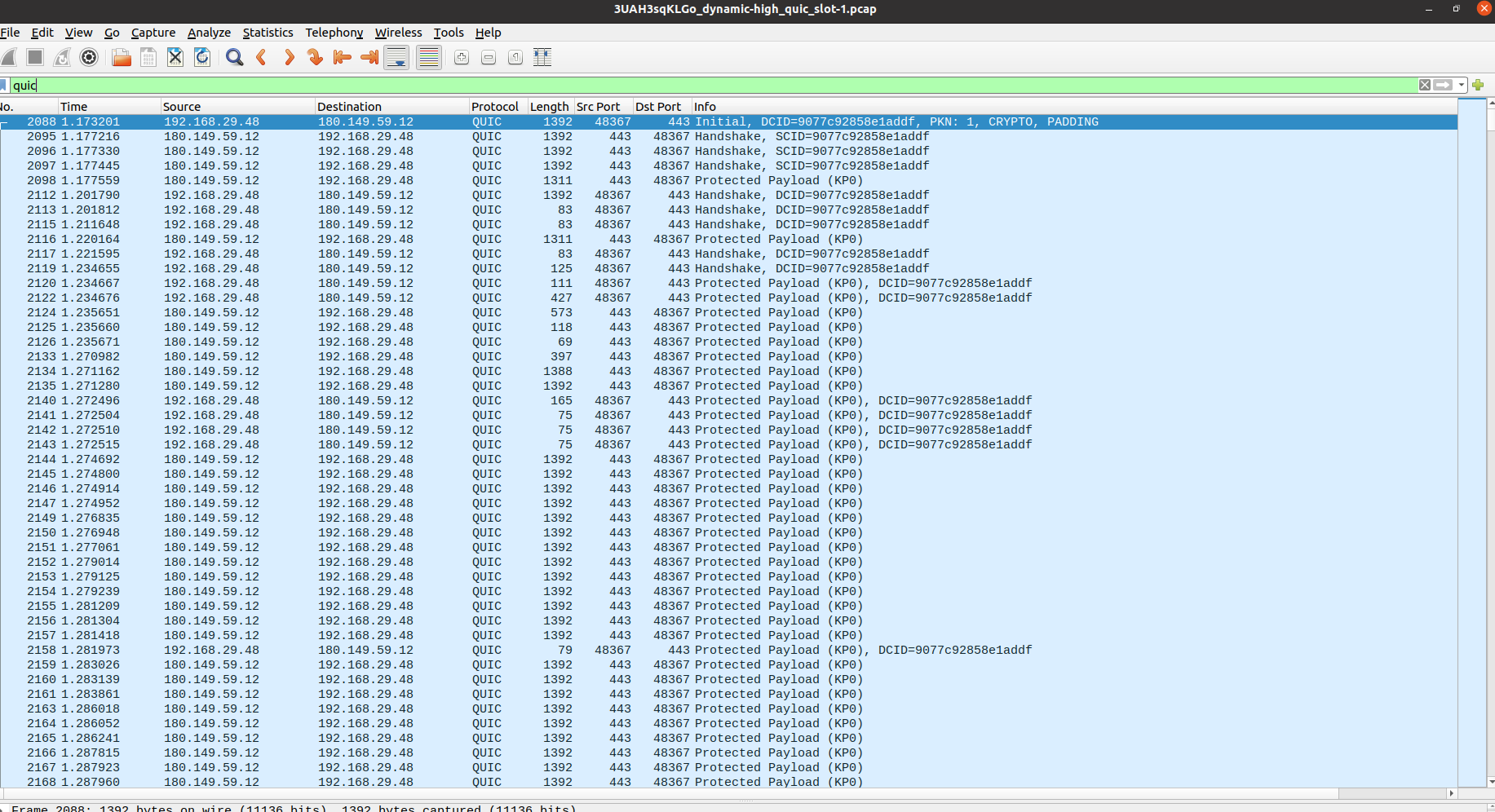}
\caption{QUIC-enabled stream at high bandwidth (DH): shows no QUIC packet for about $1.17$ sec}
\label{start-delay-high}
\end{figure}

\begin{figure}[H]
\centering
\includegraphics[width =0.95\textwidth]{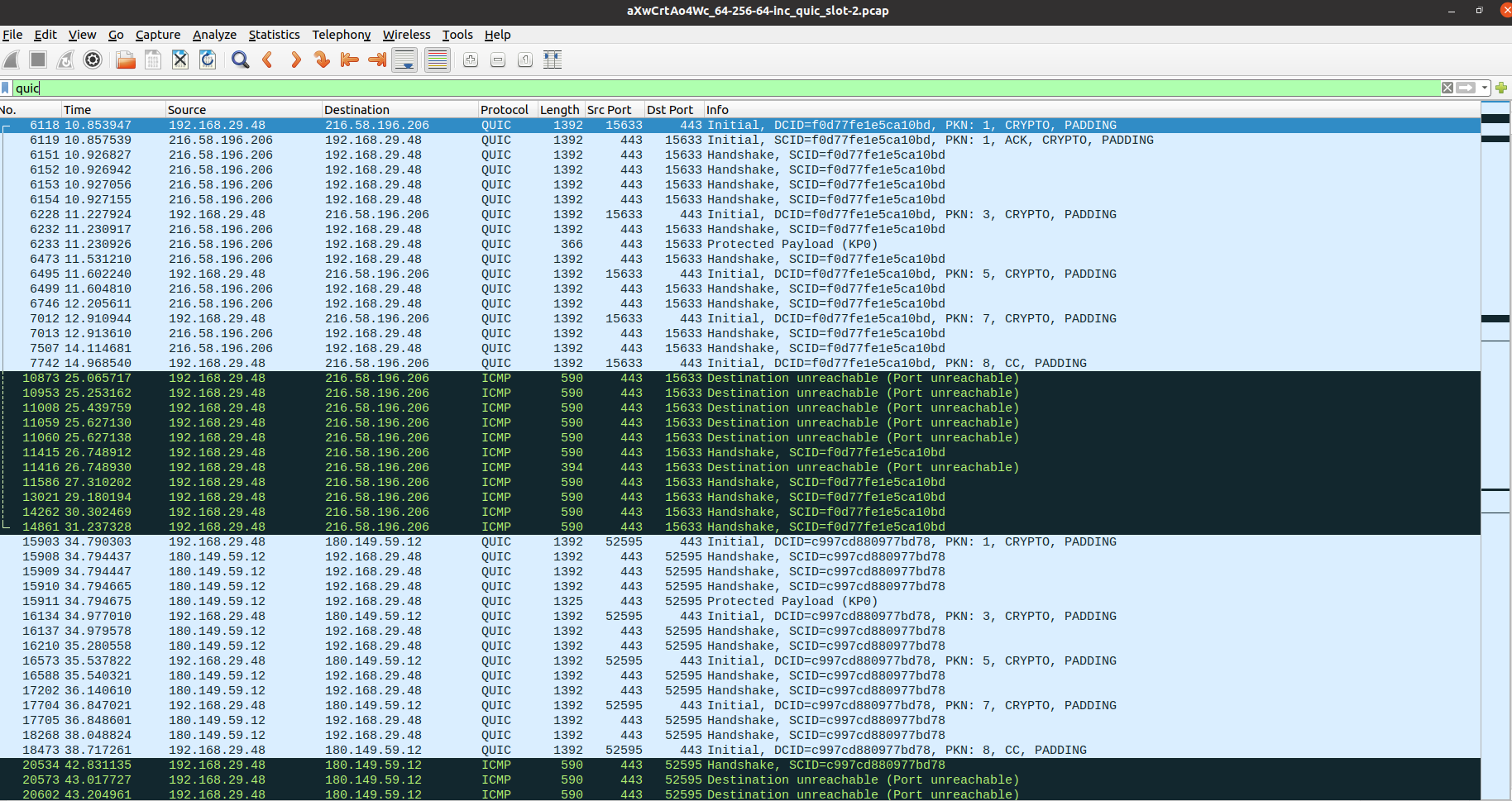}
\caption{QUIC-enabled stream at low bandwidth (DVL): shows no QUIC packet for about $10$ sec}
\label{start-delay-low}
\end{figure}

\begin{figure}[H]
\centering
\includegraphics[width =0.95\textwidth]{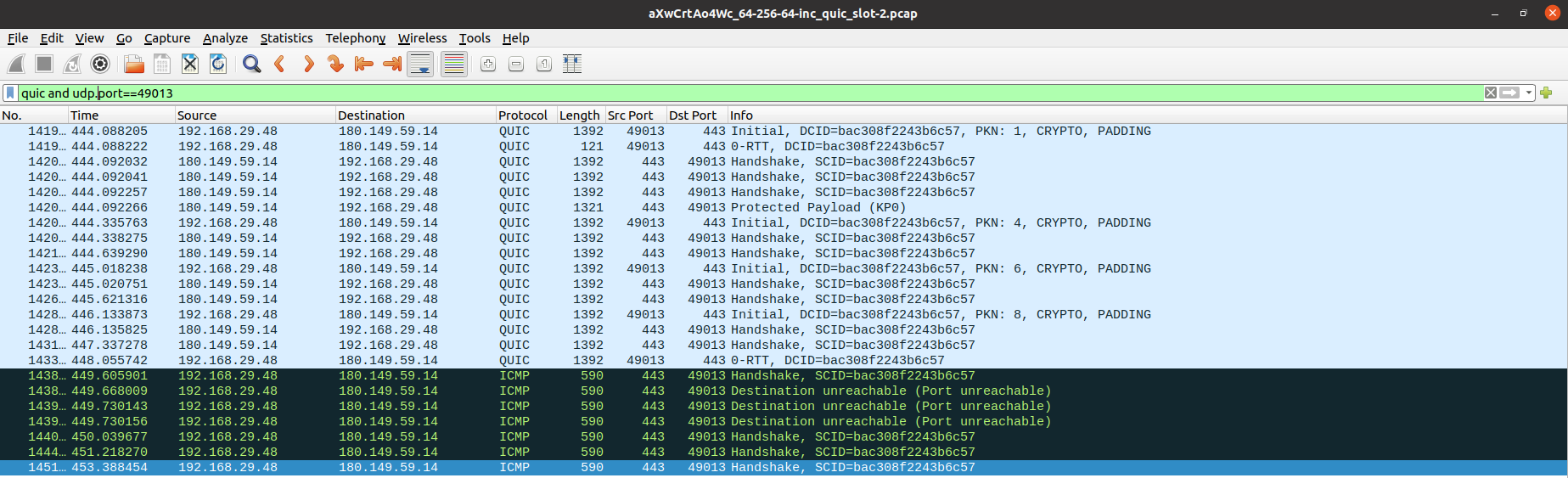}
\vspace{-0.4cm}
\caption{QUIC-enabled stream: tug of war between QUIC and TCP, shows QUIC connection-establishment fails}
\label{tug-of-war1}
\end{figure}

\begin{figure}[H]
\centering
\includegraphics[width=0.95\textwidth]{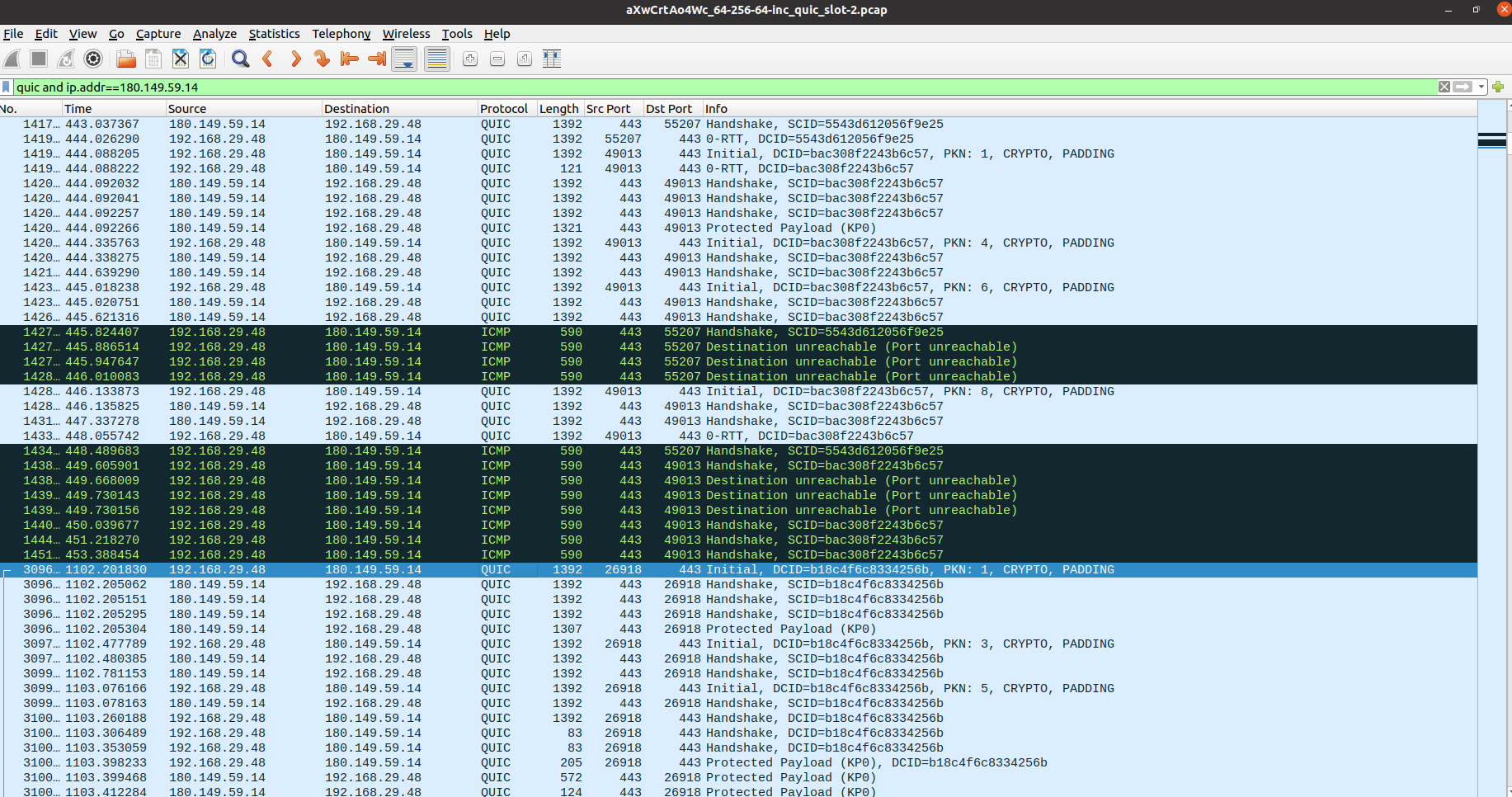}
\caption{QUIC-enabled stream: tug of war between QUIC and TCP, shows no QUIC packet to that server for $650$ sec}
\label{tug-of-war2}
\end{figure}

\begin{figure}[H]
\centering
\includegraphics[width =0.95\textwidth]{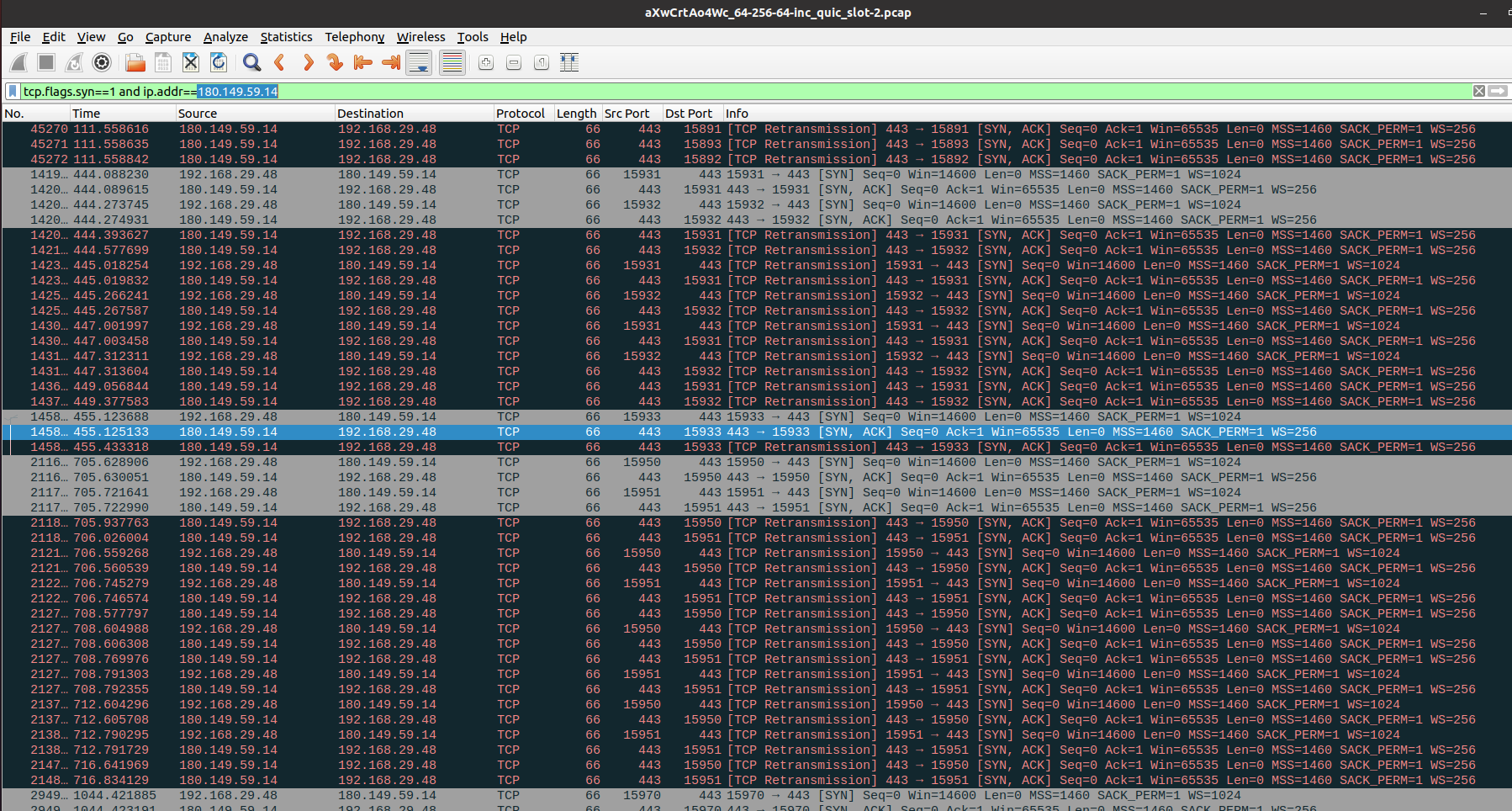}
\caption{QUIC-enabled stream: tug of war between QUIC and TCP, shows TCP connection initiated at $455$ sec with server }
\label{tug-of-war3}
\end{figure}

\begin{figure}[H]
\centering
\includegraphics[width =0.95\textwidth]{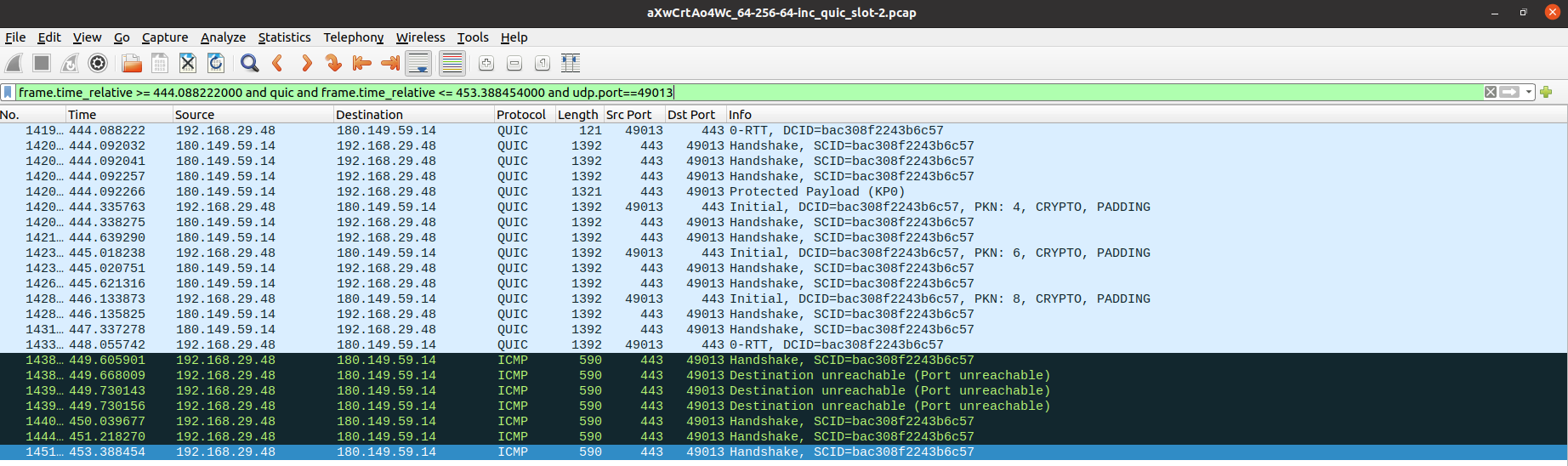}
\caption{QUIC-enabled stream: tug of war between QUIC and TCP shows 0-RTT connection fails }
\label{tug-of-war4}
\end{figure}

We conducted packet-level analysis for collected video sessions.
Fig.~\ref{start-delay-high} and Fig.~\ref{start-delay-low} show that there no QUIC packets from a QUIC enabled stream for about $1.17$ sec and $10.58$ sec for DH and DVL bandwidth respectively.
Fig.~\ref{tug-of-war1} shows an example scenario at DVL where fallback happened. We observe from Fig.~\ref{tug-of-war1} that QUIC tries to establish 1-RTT connection with the server with IP address \textit{180.149.59.14} from port number $49013$. The client sends an initial 0-RTT request; the server sends a Handshake response. Yet, probably due to packet losses, the client does not receive this. Hence, it re-initiates these requests again. This process continues $4$ times when the client's network layer finally generates an ICMP response to the server saying port unreachable. However, we immediately see another Handshake request from the client; but a timeout occurs by then. In Fig.~\ref{tug-of-war2}, we notice that the browser probably marked QUIC as broken for this connection. We do not see any packet from the client to this server for around $650$ sec. At this point in Fig.~\ref{tug-of-war3}, we notice a new TCP connection is initiated with the same server at $455$ sec. Fig.~\ref{tug-of-war4} shows that a 0-RTT connection attempt fails where the client first sends a 0-RTT request and then sends the Initial request again three times.
\end{document}